\documentclass[prb,aps,epsf,twocolumn,showpacs,10pt]{revtex4-1}
\usepackage{graphicx,amsfonts,times,bm,amsmath,verbatim,color,array}

\newcommand{\<}{\langle}

\renewcommand{\>}{\rangle}
\renewcommand{\(}{\left(}
\renewcommand{\)}{\right)}

\begin{document}
\title{Tunneling conductance for Majorana fermions in spin-orbit coupled semiconductor-superconductor heterostructures using superconducting leads}

\author{Girish Sharma}
\author{Sumanta Tewari}

\affiliation{ Department of Physics and Astronomy, Clemson University, Clemson, SC 29634}

\begin{abstract}
It has been recently pointed out that the use of a superconducting (SC) lead instead of a normal metal lead can suppress the thermal broadening effects in tunneling conductance from Majorana fermions, helping
reveal the quantized conductance of $2e^2/h$. In
this paper we discuss the specific case of tunneling conductance with SC leads of spin-orbit coupled semiconductor-superconductor (SM-SC) heterostructures in the presence of a Zeeman field, a system which has been extensively studied both theoretically and experimentally using a metallic lead. We examine the
$dI/dV$ spectra using a SC lead for different sets of physical parameters including temperature,
tunneling strength, wire length, magnetic field, and induced SC pairing potential in the SM nanowire. We conclude that in a finite wire the Majorana splitting energy $\Delta E$, which has non-trivial dependence
on these physical parameters, remains responsible for the $dI/dV$ peak broadening, even
when the temperature broadening is suppressed by the SC gap in the lead. In a finite wire the signatures of Majorana fermions with a SC lead are oscillations of quasi-Majorana peaks about bias $V=\pm\Delta_{\text{lead}}$, in contrast to the case of metallic leads where such oscillations are about zero bias. Our results will be useful for
analysis of future experiments on SM-SC heterostructures using SC leads. 
\end{abstract}

\maketitle
\section{Introduction}
In contrast to Majorana fermions (MFs) as understood in high energy physics~\cite{Perkins}, MFs in condensed matter are not elementary particles, but rather refer to collective excitations of a complex many-body ground state~\cite{Kitaev:2001, Read:2000}. However, similar to free MFs as elementary particles, these quasiparticles are also their own anti-particles, satisfying the relation $\gamma_0=\gamma_0^\dagger$, where $\gamma_0$ is the second-quantized Majorana operator. Strikingly different from ordinary Dirac fermions, MFs in condensed matter obey non-Abelian exchange statistics~\cite{Read:2000, Moore:1991, Nayak:1996, Ivanov:2001, Stern:2004}, and thus can be braided to perform fault-tolerant topological quantum computation (TQC)~\cite{Kitaev:2001, Nayak:2008}. This unconventional feature has provided an added impetus to realize MFs in a laboratory, and has resulted in an avalanche of theoretical and experimental studies~\cite{Fu:2008, Zhang:2008, Sato:2009, Tewari:2007, Sau1:2010, Tewari:2010, Sau2:2010, Lutchyn:2010, Oreg:2010, Mourik:2012, MTDeng:2012, Das:2012, Rokhinson:2012, Churchill:2013, Finck:2013, Nadj:2013, Yazdani:2014}.  

A key mechanism required for emergence of Majorana excitations in solid-state is chiral p-wave superconductivity~\cite{Kitaev:2001,Read:2000} (SC), in a low dimensional ($d\leq 2$) system of spinless (or spin-polarized) fermions. Such a mechanism supports Majorana bound states (MBS), occurring exactly at zero-energy, and localized at the defects of the order parameter in the system. Even though p-wave pairing of spinless fermions has a rather unphysical Hamiltonian, there have been a host of proposals to mimic the mean field spinless p-wave superconducting Hamiltonian in realistic systems, such as on topological insulator-superconductor interface~\cite{Fu:2008}, cold atom fermionic gases~\cite{Zhang:2008, Sato:2009}, and superconductor-semiconductor heterostructures~\cite{Tewari:2007, Sau1:2010, Tewari:2010, Sau2:2010, Lutchyn:2010, Oreg:2010}. Subsequent experiments have detected signatures of the existence of these modes in semiconductor-superconductor heterostructures~\cite{Mourik:2012, MTDeng:2012, Das:2012, Rokhinson:2012, Churchill:2013, Finck:2013}, however there has been no unique confirmation of a MBS from these experiments. Very recently MBSs were proposed~\cite{Nadj:2013, Yazdani:2014, Li:2014, Brydon:2014, Hui:2014}, and claimed to be experimentally observed~\cite{Yazdani:2014}, in Fe atomic chains embedded on a superconducting Pb [110] surface.  

Kitaev's spinless 1D p-wave chain can be physically realized in a 1D semiconductor-superconductor heterostructure nanowire~\cite{Sau2:2010, Lutchyn:2010, Oreg:2010}, in the presence of spin-orbit coupling (SOC), proximity induced superconductivity, and Zeeman splitting above a critical value $\Gamma_c$. The zero-energy Majorana modes occurring at the two ends of this 1D topological superconducting nanowire can be inferred in tunneling experiments using metallic leads, where these zero-energy modes should give rise to a peak in the differential tunneling conductance ($dI/dV$) exactly at zero bias voltage~\cite{Sau2:2010,Law:2009, Flensberg:2010}. This zero bias peak (ZBP) has been observed in experiments on semiconductor-superconductor heterostructure nanowires under appropriate physical conditions~\cite{Mourik:2012, MTDeng:2012, Das:2012, Churchill:2013, Finck:2013}. However, the presence of a ZBP alone does not provide an unambiguous signature of MFs, as even other topologically trivial subgap states may also produce a similar response~\cite{Liu:2012, Roy:2013, Pikulin:2012}. A unique `smoking-gun' signature of the Majorana ZBP is its quantized value $G=2e^2/h$ which should be observed in an ideal transport experiment from MFs. So far, this distinguishing feature has not been observed in experiments due to multiple factors (see below) which broaden the lineshape, and it thus remains an outstanding problem to reproduce the predicted peak height from putative MBS excitations from topological superconducting nanowires. 

The reduction of Majorana ZBP height from its quantized value of $2e^2/h$ in semiconductor-superconductor heterostructure nanowire is due to two principle factors: finite temperature effects, and overlapping Majorana wavefunctions from the two ends. If the temperature is larger than the tunneling strength, the ZBP is significantly broadened~\cite{Pientka:2012}. Very recently, Peng et al.~\cite{Peng:2015} has proposed to counter this problem with the use of a SC lead in place of a more commonly used metallic one. The SC lead suppresses the effects of thermal broadening because of the spectral gap ($\Delta_{\text{lead}}$ in the lead itself). With a SC lead, the Majorana peak no longer shows up at zero bias, but is shifted symmetrically to $\pm \Delta_{\text{lead}}$, with a new peak height of $G_M=(4-\pi)2e^2/h$, slightly smaller than the conventional metallic lead ZBP height $G=2e^2/h$. However, such quantization of peak height by a SC lead should be observable only from an infinitely long wire, where each Majorana mode can be manipulated individually without interference from any other low energy bound states, and especially from the other MBS localized at a different end. A more experimentally pertinent situation is, however,  that of a finite wire, where the two Majorana wavefunctions at the two  boundary points have localization lengths of the order of the wire length, and thus have a non-zero overlap. The overlap between the two MBSs moves them away from zero energy, resulting in `quasi' Majorana mode~\cite{Tudor:2013}. However, these modes are adiabatically connected with the zero energy Majorana modes in an infinitely long wire~\cite{Tudor:2013}. These effects have been theoretically investigated in the past, but in the context of a ZBP for a normal metal-SC tunneling junction~\cite{Sarma1:2012, Dumitrescu1:2015, Cheng:2009, Cheng:2010, Lin:2012}. The fate of $dI/dV$ spectra in semiconductor-superconductor heterostructure nanowires with SC lead in experimentally relevant situations and for short wires remains unexplored, and we address this important issue in this article. 

In Sec. II, we describe the tight binding formalism of topological superconductivity in a 1D nanowire and discuss its relevant symmetries. We also introduce the Hamiltonian for the SC lead, and the tunneling Hamiltonian which couples it to the substrate. In Sec. III, following Ref.~\onlinecite{Peng:2015}, we present the Green's function formalism, which is used to obtain the $dI/dV$ spectra for a SC-nanowire junction. We then present our numerical results, varying several physical parameters, such as temperature, tunneling strength, and the wire length. We examine at length the $dI/dV$ spectra using a superconducting lead for different sets of physical parameters which include temperature, tunneling strength at the junction, wire length, magnetic field, and induced superconducting pairing potential in the semiconductor nanowire. We conclude that the Majorana splitting energy $\Delta E$, which has non-trivial dependence on these physical parameters, remains primarily responsible for the $dI/dV$ peak broadening, even when temperature broadening is suppressed by the use of a superconducting lead. Our results will be useful for the analysis of future experiments on semiconductor-superconductor heterostructure using superconducting leads. We conclude with a summary and discussion in Sec. IV.

\section{Model Hamiltonian}
The non-interacting single particle Hamiltonian for a spin-orbit coupled nanowire subjected to to Zeeman field ($h_x$) can be written as
\begin{align}
H_0 = -\frac{\partial_x^2}{2m*} - \mu + h_x\sigma_x + i\alpha\partial_x\sigma_y,
\end{align}
where $m*$, $\alpha$, and $\mu$ are the effective mass, spin-orbit strength, and chemical potential respectively. The proximity induced superconductivity (with mean field strength $\Delta$) can be described by $H_\Delta = \Delta (c^\dagger_{\uparrow}(x)c^\dagger_{\downarrow}(x))+$ h.c. The mean field BCS superconducting Hamiltonian will be given by $H_{BCS} = H_0 + H_\Delta$. Quasiparticle excitations above this many-body ground state are described by the Bogoliubov-de Gennes equation $H_{\text{BdG}}\Psi=E\Psi$, where $H_{\text{BdG}}$ is the Bogoliubov-de Gennes Hamiltonian constructed as 
\begin{align}
H_{\text{BdG}} = \left( \begin{array}{cc}
H_0 & \Delta(x)   \\
\Delta^*(x) & -\sigma_y H^*_0\sigma_y  \\
\end{array} \right),
\end{align}
written in the Nambu basis $\Psi=[u_{\uparrow}(x),u_{\downarrow}(x),v_{\downarrow}(x),-v_{\uparrow}(x)]^T$. Non-trivial Majorana modes, which are zero-energy excitations of the superconducting many-body ground state, are given by the condition $H_{\text{BdG}} \Psi = 0$, which emerge in the topological superconducting phase of the Hamiltonian, satisfying the relation $h_x>\sqrt{\mu^2+\Delta^2}$. We will now study the electronic tight-binding description of this Hamiltonian, and examine its relevant symmetries. We will then introduce the SC lead Hamiltonian $H_{\text{lead}}$ which couples to the substrate Hamiltonian at the chain end $x=0$, via the tunneling Hamiltonian $H_{T}(\tau)$.

The tight binding Hamiltonian for a one-dimensional spin-orbit coupled nanowire, with proximity induced superconductivity, and a magnetic field can be written as
\begin{align}
H &= \sum\limits_{r=1}^N [-\mu (c^\dagger_{r\uparrow} c_{r\uparrow} + c^\dagger_{r\downarrow} c_{r\downarrow}) + \Delta (c^\dagger_{r\uparrow}c^\dagger_{r\downarrow} - c^\dagger_{r\downarrow}c^\dagger_{r\uparrow}) \nonumber\\
&+ h_x (c^\dagger_{r\uparrow} c_{r\downarrow} +  c^\dagger_{r\downarrow} c_{r\uparrow})-t(c^\dagger_{r\uparrow} c_{r+1\uparrow} + c^\dagger_{r\downarrow} c_{r+1\downarrow})\nonumber\\
&+\alpha(c^\dagger_{r\uparrow} c_{r+1\downarrow} - c^\dagger_{r\downarrow} c_{r+1\uparrow})]+ \mbox{h.c},
\label{Eq_H_1}
\end{align}
where the sum $r$ is over all the lattice sites with open boundary conditions at $r=1$, and $r=N$, where $N$ is the number of lattice sites. In Eq.~\ref{Eq_H_1}, $\mu$ represents the chemical potential, $\Delta$ is the proximity induced $s-$wave superconducting pairing potential, $h_x$ is the magnetic field applied in the $x$ direction, which is also assumed to be the direction of the wire, $t$ is the hopping integral for the nearest neighbor site on the nanowire, and $\alpha$ is the spin-orbit coupling strength. The operator $c_{r,s}$ $(c^\dagger_{r,s})$ annihilate (create) an electron on the lattice site $r$ with spin $s=\uparrow$ or $\downarrow$. When the magnetic field $h_x$ exceeds a critical value $h_c$, the Hamiltonian in Eq.~\ref{Eq_H_1} describes the topological superconducting phase of the SOC coupled nanowire, supporting zero-energy topologically protected Majorana modes at the boundary points.  One can define the electron annihilation operator in the momentum space as $ c_{j,s} = \frac{1}{\sqrt{N}}\sum_{k}{e^{ijk}c_{k,s}} $, where $k\equiv k_x$, $j$ represents the site position, and $c^\dagger_{k,s}$ creates an electron with momentum $k$ and spin $s$. Combining this with Eq.~\ref{Eq_H_1}, the Hamiltonian $H$ written in terms of the Fourier transformed operators $c_{k,s}$ and $c^\dagger_{k,s}$ becomes
\begin{align}
H &= \sum_k [(-t\cos k(c^\dagger_{k\uparrow}c_{k\uparrow} + c^\dagger_{k\downarrow}c_{k\downarrow})+2\Delta c^\dagger_{k\uparrow}c^\dagger_{-k\downarrow}\nonumber\\
&-\mu(c^\dagger_{k\uparrow}c_{k\uparrow}+c^\dagger_{k\downarrow}c_{k\downarrow})+h_x(c^\dagger_{k\uparrow}c_{k\downarrow}+c^\dagger_{k\downarrow}c_{k\uparrow})\nonumber\\
&-2i\alpha\sin k c^\dagger_{k\downarrow}c_{k\uparrow}]+\mbox{h.c.}
\label{Eq_Hk_1}
\end{align}
The corresponding momentum space Bogoliubov-de Gennes (BdG) Hamiltonian $H_{\text{BDG}}$,  written in the Nambu basis $\Psi_k=(c_{k\uparrow}, c_{k\downarrow}, c^\dagger_{-k\downarrow}, -c^\dagger_{-k\uparrow})^T$  is the following
\begin{align}
&H_{\text{BDG}} = 2\sum\limits_k {\Psi_k^\dagger H_k \Psi_k}, \text{where} \nonumber \\
&H_k = -(t\cos k+\mu)\sigma_0\tau_z + h_x \sigma_x\tau_0 \nonumber\\
& - \alpha \sin k \sigma_y\tau_z + \frac{\Delta}{2} \sigma_0 \tau_x,
\label{Eq_H_BDG_1}
\end{align}
assuming the order parameter $\Delta$ to be real. The identity matrices $\sigma_0$ and $\tau_0$ act in spin and particle-hole space respectively. 

The BdG Hamiltonian $H_{\text{BDG}}$ in Eq.~\ref{Eq_H_BDG_1} is symmetric under particle-hole (PH) transformation, and satisfies the following reality condition in momentum space: $\Xi H_k \Xi^{-1} = -H_{-k}$, where $\Xi=\sigma_y\tau_y\mathcal{K}$ is the anti-unitary PH operator in our chosen Nambu basis  ($\mathcal{K}$ denoting complex conjugation). The Hamiltonian $H_{\text{BDG}}$ also admits a chiral symmetry $\mathcal{S}$~\cite{Tewari1:2012}. The operator $\mathcal{S}$ can be obtained by first identifying another operator $\mathcal{O}=\mathcal{K}$ with $\mathcal{O}^2 = +1$, which acts on the Hamiltonian like a pseudo time-reversal operator satisfying $\mathcal{O}H_k\mathcal{O}^{-1} = H_{-k}$. The chiral operator $\mathcal{S}$ is then just a product of PH and pseudo TR operator: $\mathcal{S}=\mathcal{O}\cdot\Xi$. The operator $\mathcal{S}=\sigma_y\tau_y$ anti-commutes with $H_{\text{BDG}}$ i.e. $\{H_{\text{BDG}}, \mathcal{S}\}=0$, and thus the Hamiltonian $H_{\text{BdG}}$ belongs to chiral class BDI characterized by an integer $\mathbb{Z}$ invariant.

The BCS Hamiltonian for the SC lead, in momentum space, can be written as
\begin{align}
H_{\text{lead}} = \sum_{\mathbf{k},\sigma}{\zeta_{\mathbf{k}}c^\dagger_{L,\mathbf{k},\sigma}c_{L,\mathbf{k},\sigma}+(\Delta_{\text{lead}} c_{L,\mathbf{k},\uparrow}c_{L,-\mathbf{k},\downarrow}+\text{h.c})},
\label{Eq_H_lead_1}
\end{align}
where $\Delta_{\text{lead}}$ is the SC gap, and the operator $c_{L,\mathbf{k},\sigma}$ annihilates an electron with spin $\sigma$ on the lead, and $\zeta_{\mathbf{k}}=k^2/2m-\mu_{\text{lead}}$. The tunneling strength between the lead (Eq.~\ref{Eq_H_lead_1}) and the substrate (Eq.~\ref{Eq_H_1}) is represented by the hopping integral $t'$, and the tunneling Hamiltonian is given by
\begin{align}
H_T(\tau) = \sum_{\sigma}{t' e^{i\phi(\tau)/2} c^\dagger_{L,\sigma}(0,\tau)c_{\sigma}(0,\tau)} + \text{h.c},
\label{Eq_H_T}
\end{align}
where $\tau$ is the time argument. The real space operator $c^\dagger_{L,\sigma}(x,\tau)$ is the Fourier transform of the operator $c^\dagger_{L,\mathbf{k},\sigma}$ in Eq.~\ref{Eq_H_lead_1}. The substrate and the lead are in contact at $x=0$, which is the argument of the electron operator on the lead $c^\dagger_{L,\sigma}(0,\tau)$ and on the substrate $c_{\sigma}(0,\tau)$. The phase difference between the lead and the sample is given by $\phi(\tau)$. 

In our numerical analysis, we will focus only on the weak tunneling regime, which is given by the condition $\omega_t\ll \Delta_{\text{lead}}$~\cite{Peng:2015}, where $\omega_t = (\pi t'^2 \nu_0 |\zeta(0)|^2\sqrt{\Delta_{\text{lead}}/2})^{2/3}$. The quantity $|\zeta(0)|^2 = |u_\uparrow|^2 + |u_\downarrow|^2$ for the Majorana Nambu spinor $[u_\uparrow, u^*_\downarrow, u_\downarrow, -u^*_\uparrow]^T$ at $x=0$, and $\nu_0$ is the normal density of states at the Fermi energy in the SC lead. From our numerical estimate of the Majorana wavefunctions, we find that $|\zeta(0)|^2\sim 0.001$, and thus choosing $\nu_0=1$, $\Delta_{\text{lead}}\sim\Delta\sim 500 \mu eV$, suggests that choosing $t'\sim 20\mu eV$ satisfies the weak tunneling condition. For our numerical analysis in the next section, we will choose our parameter values such that the tunneling between the SC lead and the substrate is weak. Further, for all our numerical results in this paper, we use the values of physical parameters roughly consistent with the properties of InSb nanowires~\cite{Mourik:2012}. We chose the effective mass $m^*=0.015m_e$, Rashba SOC strength $\alpha=0.2 meV$, hopping integral $t=\hbar^2/2m^*a^2$, lattice spacing $a=15 nm$, and $\mu=-2t$, fixed throughout. We will vary the SC gap $\Delta$ and the applied magnetic field $h_x$, both in the range $0.5 meV-$ $1 meV$.  

\section{Differential Tunneling conductance with Superconducting lead}
In the context of semiconductor-superconductor heterostructures in condensed matter, the Majorana Fermion manifests as a sub-gap zero-energy mode. A simple method to verify the existence of this exotic mode is through the detection of a zero-bias peak in the tunneling conductance measurement between a metallic lead and the topological superconductor~\cite{Law:2009, Flensberg:2010}. More importantly, the ZBP is characterized by its quantized peak height $G=2e^2/h$~\cite{Law:2009, Flensberg:2010}. Even though signatures of Majoranas in terms of zero-bias peaks have been observed in a series of recent experiments~\cite{Mourik:2012, Das:2012, Churchill:2013, MTDeng:2012, Finck:2013, Yazdani:2014}, the observation of the predicted quantized theoretical value remains a pressing issue. This has been primarily attributed to the effects of peak broadening at finite temperature and overlap due to shorter wire lengths. Therefore, uniquely distinguishing these peaks from other  possible non-topological zero-energy states sub-gap states~\cite{Liu:2012, Roy:2013, Pikulin:2012} is challenging. In this section we will study the $dI/dV$ characteristics of a superconductor-semiconductor heterostructure nanowire, using a SC lead as a conductance probe.

Using the method of superconducting lead as a conductance probe, the tunneling current $I$ using superconducting lead (with gap $\Delta_{\text{lead}}$) is given by~\cite{Peng:2015}
\begin{align}
I(V)=4e\pi^2{t'}^4\int\frac{d\omega}{h}[&\mbox{Tr}(g_{eh}(r,\omega)g^\dagger_{eh}(r,\omega))\rho(\omega_-)\rho(\omega_+)\nonumber\\
&(n_F(\omega_-)-n_F(\omega_+))],
\label{Eq_I_V_1}
\end{align}
where $t'$ denotes the tunneling strength between the sample and the superconducting lead, $\omega_{\pm}=\omega\pm eV$, $n_F(\omega)$ represents the Fermi-Dirac distribution function $(e^{\omega/T}+1)^{-1}$, $\rho(\omega)$ is the density of states in the superconducting lead: $\rho(\omega)=\nu_0 \theta(|\omega|-\Delta)|\omega|/\sqrt{\omega^2-\Delta_{lead}^2}$, and $g_{eh}(r,\omega)$ is the retarded Green's function in the electron-hole subspace. Now the Majorana Fermion peaks no longer appear at zero bias, but are shifted by the superconducting gap $\Delta_{lead}$ to $V=\pm\Delta_{lead}$. Secondly the peak is asymmetric around $V/\Delta_{lead}=\pm 1$, and sharply rises at the threshold $\Delta_{lead}$ (see Figure~\ref{dIdV_temperature}). The theoretical peak height in this case is: $G_M=\max (dI/dV) = (4-\pi)2e^2/h\approx 1.72 e^2/h$~\cite{Peng:2015}, slightly smaller than the quantum of electrical conductance $G=2e^2/h$.

Once the real space Hamiltonian of the substrate is defined (see Eq.~\ref{Eq_H_1}), it is then possible to obtain the Green's function for the system as:
\begin{align}
G_0(\omega) = [(\omega+i\epsilon)I - H]^{-1}=\sum\limits_m\frac{|\phi_m\>\<\phi_m|}{\omega-E_m+i\epsilon},
\label{Eq_Green_1}
\end{align}
where $E_m$ is an energy eigenvalue of Hamiltonian, with corresponding eigenstate $|\phi_m\rangle$, and $\epsilon$ is positive infinitesimal. $G_0(\omega)$ in Eq.~\ref{Eq_Green_1} contains all the degress of freedom, namely spatial, spin, and particle-hole, making it a $4N$ dimensional object for $N$ lattice sites. The local Green's function at a coordinate $r$ is given by: $g^0_r(\omega)=G(r,\omega)$, which is a four component matrix in the Nambu space, for a specific position $r$ in the one-dimensional chain. However $G_0(\omega)$ (or even $g^0_r(\omega)$) is entirely for the Hamiltonian $H$ in Eq.~\ref{Eq_H_1}, which does not take into account the tunneling between the SC lead and the nanowire substrate. 

The BCS Hamiltonian for the SC lead was introduced in Eq.~\ref{Eq_H_lead_1} of the previous section. The tunneling strength between the lead (Eq.~\ref{Eq_H_lead_1}) and the substrate (Eq.~\ref{Eq_H_1}) is represented by the hopping integral $t'$, and given by the tunneling Hamiltonian in Eq.~\ref{Eq_H_T}. Neglecting the Andreev reflection in the lead near $eV\sim \Delta_{\text{lead}}$, the coupling between the lead and the substrate can be captured by the following self-energy term $\Sigma$~\cite{Peng:2015}
\begin{align}
\Sigma=i\pi(t')^2\left( \begin{array}{cccc}
\rho(\omega_-) & 0 & 0 &  0  \\
0 & \rho(\omega_-) & 0 & 0  \\
0 & 0 & \rho(\omega_+) & 0    \\
0 & 0 & 0 & \rho(\omega_+)   \\
\end{array} \right)
\label{Eq_Sigma}
\end{align}
The Green's function $g_r$ at a specific position $r$, supplemented by the self-energy term, can be thus obtained as: $g_r=((g^0_r)^{-1}-\Sigma^{-1})^{-1}$. Choosing $r=0$ or $r=L$ and substituting the $e$-$h$ block of $g_r$ in Eq.~\ref{Eq_I_V_1} (i.e. $g_{eh}(r,\omega)$) gives the tunneling current $I(V)$ contribution which is localized at $r=0$ and $r=L$. In the regime of weak tunneling, $dI/dV$ has an approximate analytic form for the peak lineshape which can be written as~\cite{Peng:2015}
\begin{align}
\frac{dI}{dV} = (4-\pi)\frac{2e^2}{h}\Lambda\(\frac{eV-\Delta_{\text{lead}}}{\omega_t}\),
\label{Eq_dIdV}
\end{align}
with a maximum peak height of $(4-\pi)2e^2/h$. The functions $\Lambda(z)$ and $\omega_t$ are such that when $eV\geq\Delta_{\text{lead}}$, $\Lambda(z)=1$, and when $eV<\Delta_{\text{lead}}$, $\Lambda(z)=0$ ~\cite{Peng:2015}, therefore $dI/dV$ sharply rises at $eV=\Delta_{\text{lead}}$.  In our numerical study done on a lattice, instead of using the approximate analytic form given in Eq.~\ref{Eq_dIdV}, we directly calculate $g_r(\omega)$ from Eq.~\ref{Eq_H_1} and Eq.~\ref{Eq_H_lead_1}, and use Eq.~\ref{Eq_I_V_1} to calculate $dI/dV$ for semiconductor-superconductor heterostructure nanowire  as a function of various physical parameters like $T$, $\Delta$, $\mu$, $h_x$, and $L$.

\subsection{Finite temperature effects}
\begin{figure}[h]
\centering
\includegraphics[scale=0.2]{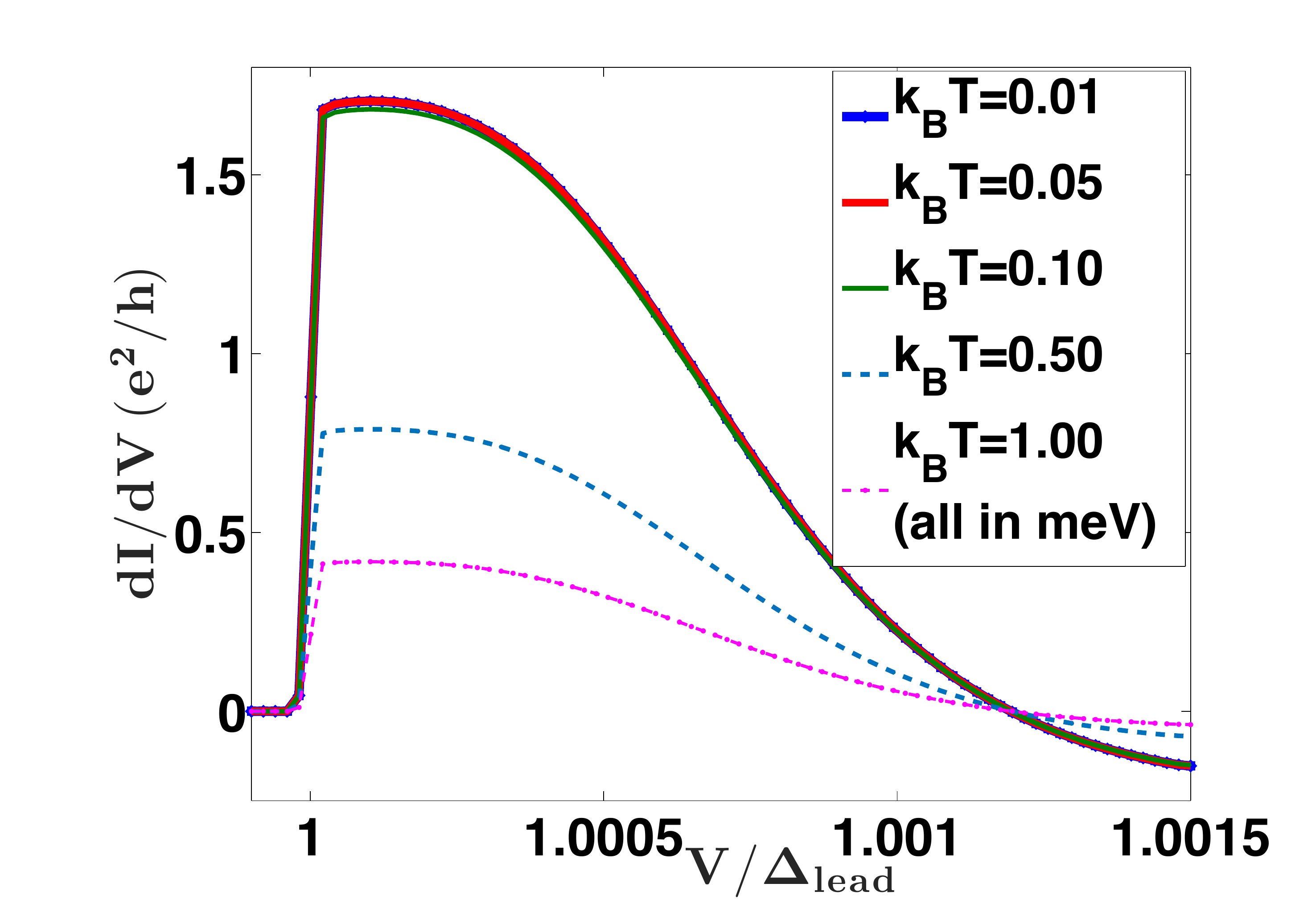}
\includegraphics[scale=0.23]{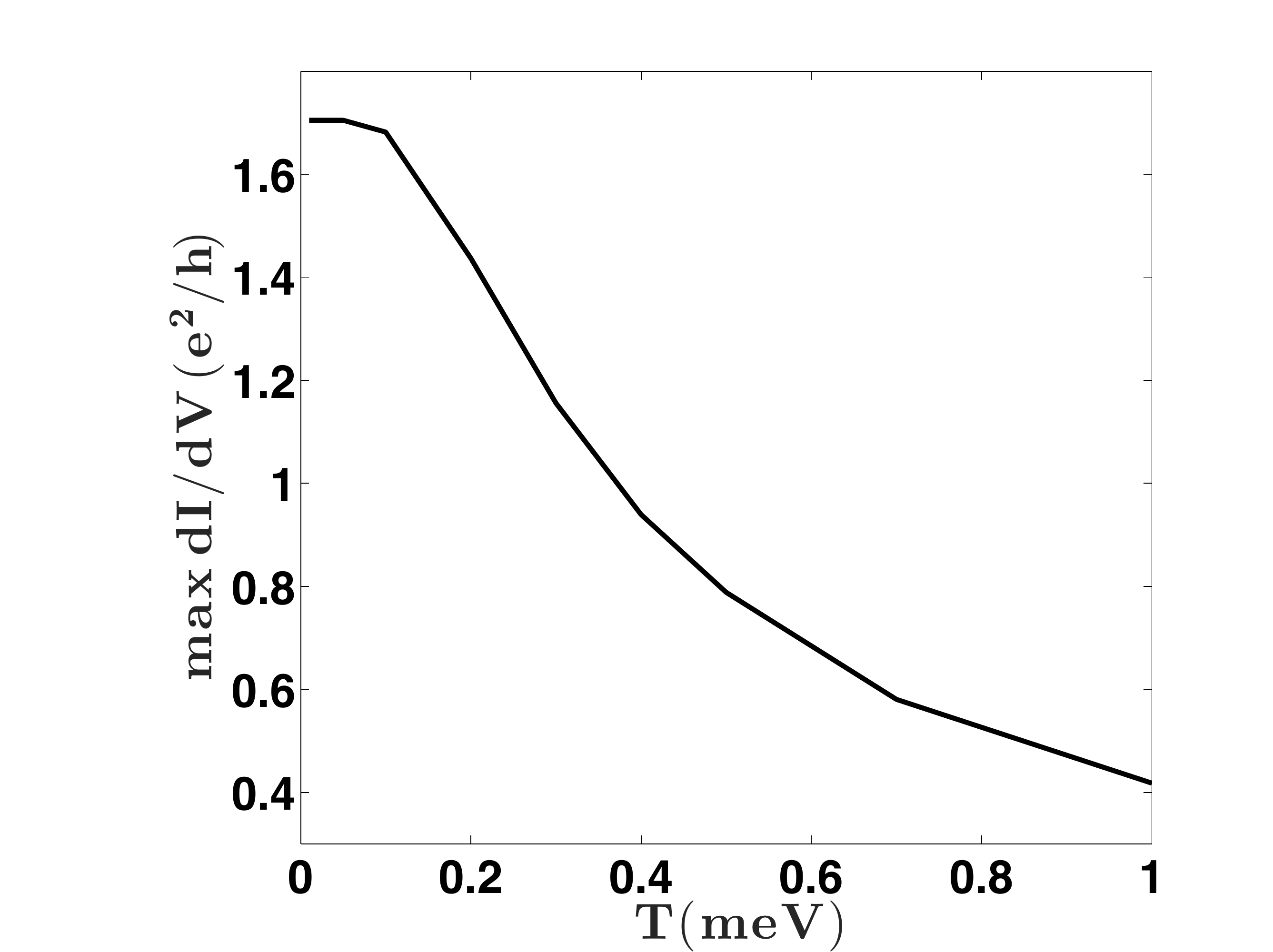}
\caption{(color online) \textit{Top panel:} Differential conductance $dI/dV$ at various temperatures (in $meV$) as a function $V/\Delta_{\text{lead}}$, where $V$ is the bias voltage and $\Delta_{\text{lead}}$ is the superconducting gap of the lead. \textit{Bottom panel:} Maximum peak height (denoted by $\max dI/dV$) as a function of temperature. The parameters used are: $L=4.2 \mu m$, $\Delta=0.5 meV$, $h_x=0.60 meV$, $\Delta_{\text{lead}}=\Delta$, $t'=25 \mu eV$.}
\label{dIdV_temperature}
\end{figure}

We now present numerical results for tunneling conductance for different sets of physical parameters. Figure~\ref{dIdV_temperature} shows the plot of $dI/dV$ as a function of $V/\Delta_{\text{lead}}$ at various temperatures. Note that in this case, in order to examine the temperature dependence solely, we work in the parameter regime where the wave-function overlap (and consequently the splitting energy $\Delta E$) is small. As expected, there is a peak at $V=\Delta_{\text{lead}}$ (see Figure~\ref{dIdV_temperature}), and another symmetrically placed one at $V=-\Delta_{\text{lead}}$ (not shown in the figure). As seen from Figure~\ref{dIdV_temperature}, at $T\sim 0$, when the Fermi function $n_F(\omega)$ reduces to $\theta(\mu-\omega)$, the peak height is quantized to its maximum theoretical value $G_M$. The maximum peak height decreases as $T$ is increased. However when compared to the use of normal metallic leads, these thermal effects are suppressed with the use of superconducting leads $\sim\exp(-\Delta_{lead}/T)$, at least in the limit $t'<T$. We note from Figure~\ref{dIdV_temperature} (right panel) that the conductance peak reduces to about $0.4G_M$ when $T$ is as high as $1 meV$. The effect of temperature broadening on the Majorana peak is therefore suppressed by the use of the SC lead in the limit when tunneling between the lead and substrate ($t')$ is much smaller than the temperature ($t'<T$). However we point out that in Fig~\ref{dIdV_temperature}, we have not considered the temperature dependence of the superconducting gap. The temperature dependence of the superconducting gap within the BCS theory is given by
$\Delta(T) \sim \Delta(T=0) \sqrt{1-T/T_C}$, where $\Delta(T=0) \sim 1.764 k_B T_C$
With increase in temperature, the superconducting gap is suppressed. When $T\sim T_C$, the gap becomes zero, and the calculation breaks down. Hence, including the temperature dependence of the gap will further suppress the peak height at a finite temperature. When  $T \sim T_C$, including this temperature dependence of the superconducting gap the  peak height will be suppressed even more. However when $T\ll T_C$ this dependence can be ignored, and the peak height stays roughly the same. Experimentally, (for example see Ref.~\onlinecite{Mourik:2012} ), the temperature range considered is $\sim 5\mu eV-30 \mu eV$, and the induced SC gap is $\sim 250 \mu eV$, which very well falls in the range $T=0.02 \Delta$ -- $0.10 \Delta$, and higher temperatures are not relevant.

Next we will examine the zero temperature $dI/dV$ profile, but in the regime when wavefunction overlap effects are not negligible. 
\subsection{Effects of wavefunction overlap}
In a one dimensional nanowire, the Majorana `zero-energy' mode occurs exactly at zero energy only in the idealized situation of an infinitely long wire. Any realistic experiment is however done at a finite non-zero temperature for wires of a finite length. For a finite wire length the two Majorana wave functions at the ends of the wire are no longer localized at the two ends but can overlap with each other. 

\begin{figure}[h]
\centering
\includegraphics[scale=0.28]{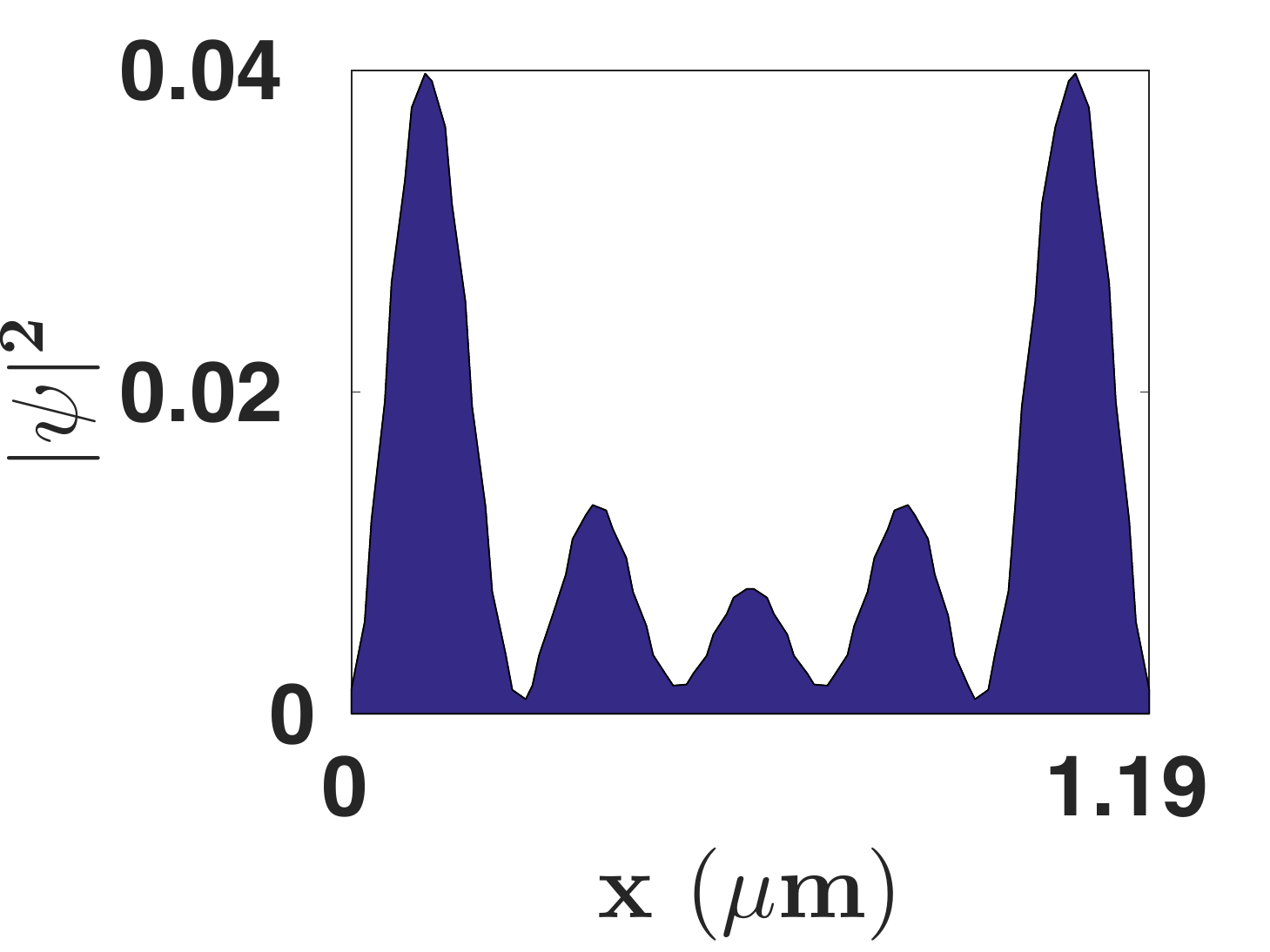}
\includegraphics[scale=0.28]{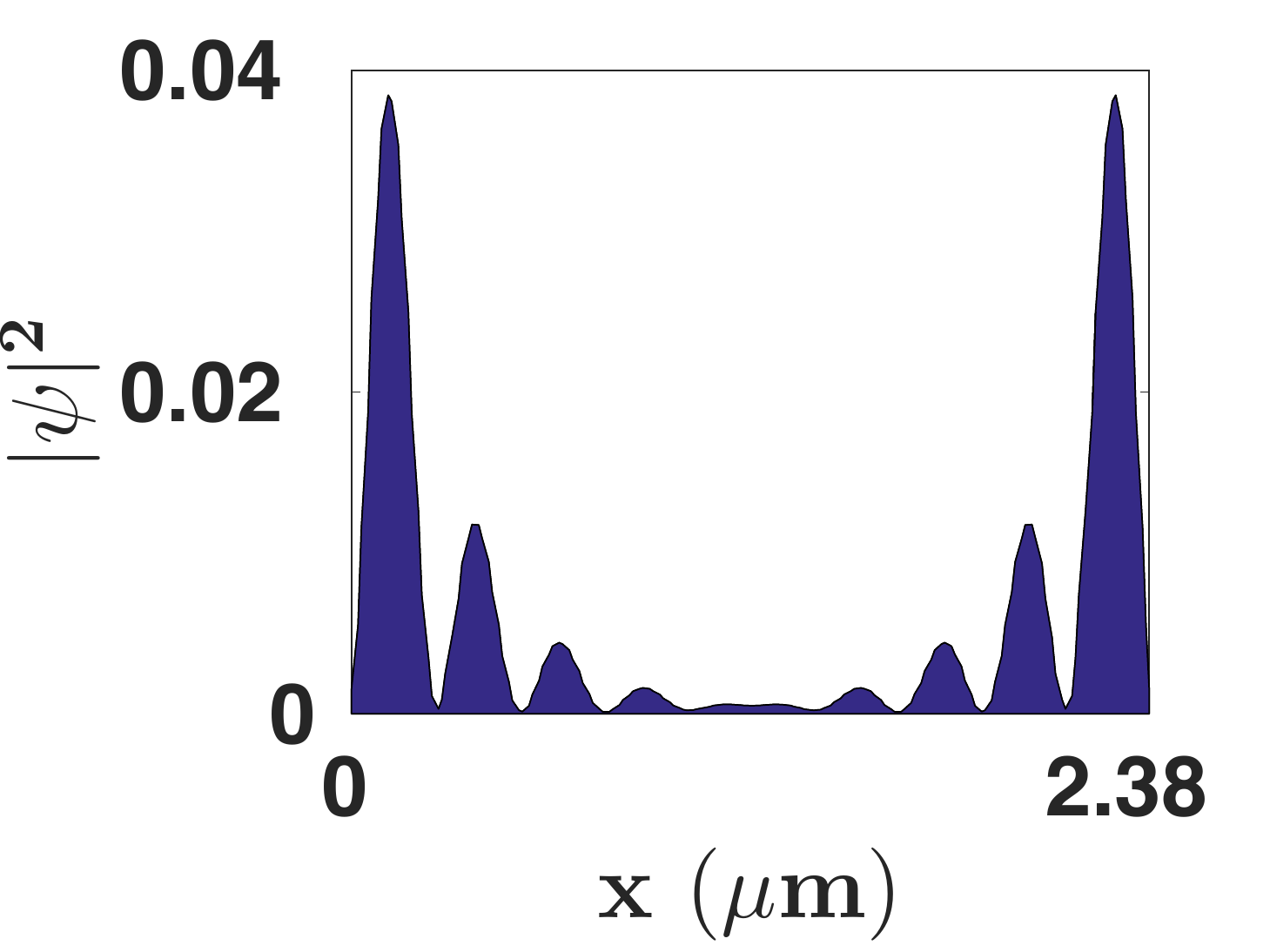}\\
\includegraphics[scale=0.28]{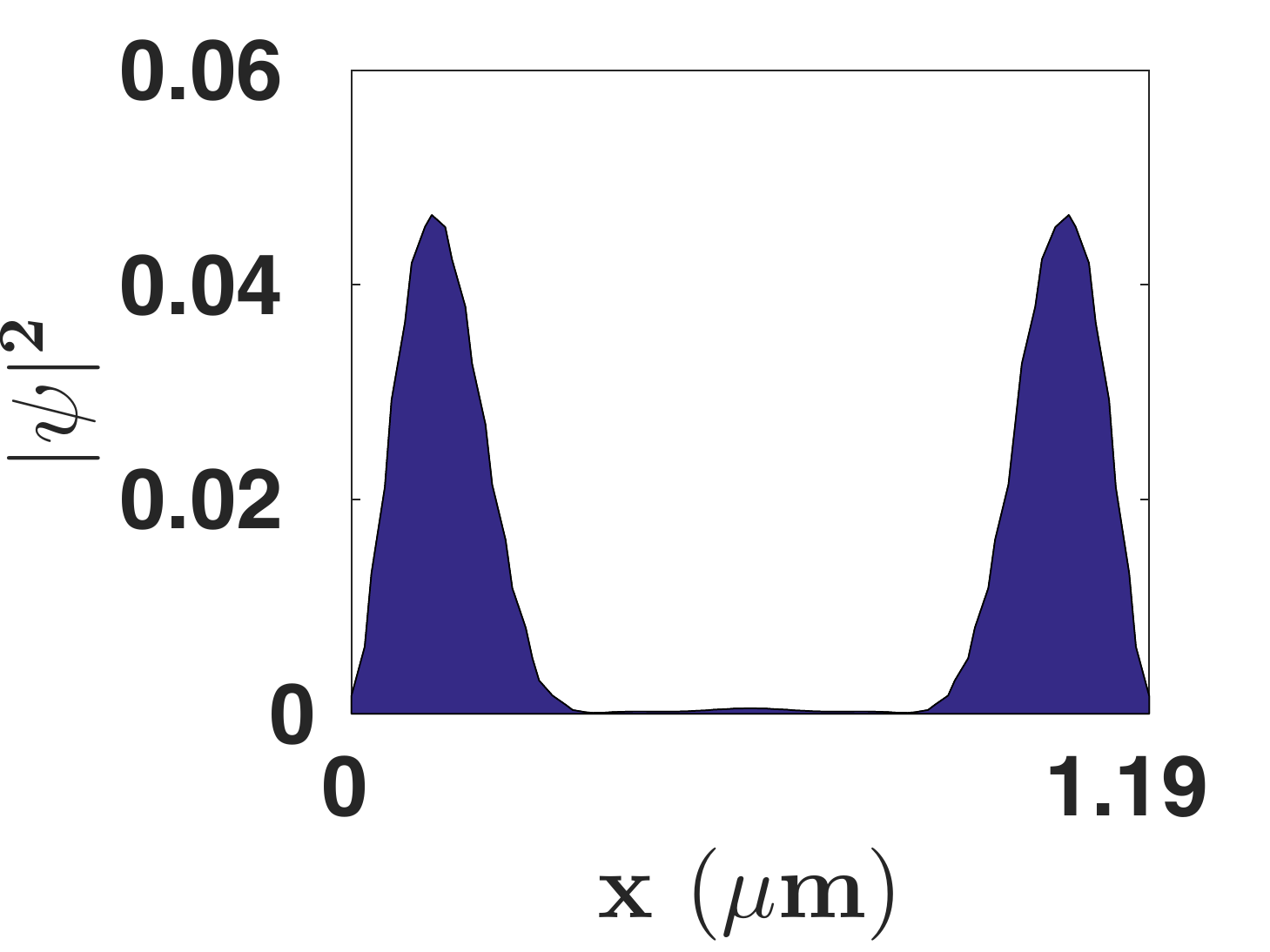}
\includegraphics[scale=0.28]{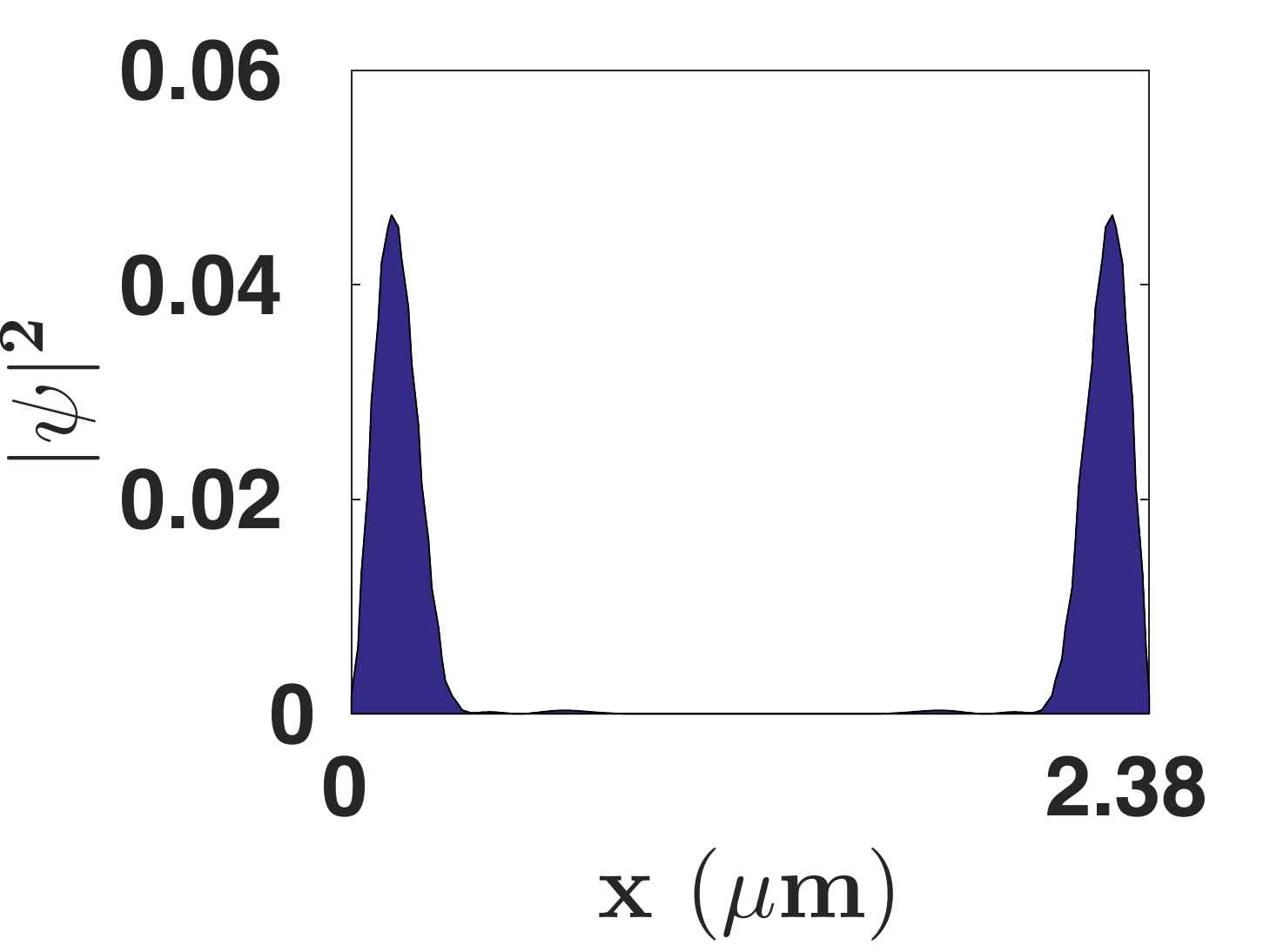}
\caption{The Majorana mode wavefunction plotted over the entire chain length for two different wire lengths $L=1.2 \mu m$ and $L=2.4 \mu m$ (left and right panels respectively), and for two different parameter sets. \textit{Top panel:} $h_x=0.6 meV$, $\Delta=0.5 meV$. \textit{Bottom panel:} $h_x=1.03 meV$, $\Delta=1 meV$. A stronger overlap is observed for the plots in top panel as a result of  a lesser proximity induced gap $\Delta$ (which increases the coherence length $\zeta$). }
\label{Figure_MF_wavefunc_1}
\end{figure}

The Majorana wave function (upto an overall normalization factor) in the TS phase of a 1D nanowire can be written as~\cite{Sarma1:2012}
\begin{align}
\Psi_{MF}(x)\propto e^{-x/\zeta}e^{ik_{F\text{eff}}x} |u\rangle,
\label{Eq_Psi_MF}
\end{align}
where $\zeta$ is the effective coherence length and $k_{F\text{eff}}$ is the effective Fermi wave-vector associated with the localized Majorana modes. Also $x$ can be measured from one of the two ends of the wire (0 or L) to represent the wavefunction for each mode. In Eq.~\ref{Eq_Psi_MF}, $|u\rangle$ represents the 4-component Nambu spinor of the wavefunction. $\Psi_{MF}(x)$ consists of an exponentially decaying factor $e^{-x/\zeta}$ which effectively binds the Majorana modes at the the boundary points, as long as the length of the nanowire $L\gg\zeta$. The Majorana wave-function decays on a length scale of  $\zeta$, which is the effective coherence length, and also consists of an oscillatory part $e^{ik_{F\text{eff}}x}$. The factors $\zeta$ and $k_{F\text{eff}}$ depend on the microscopic parameters $\Delta$, $\mu$, $h_x$, $\alpha$ of the Hamiltonian $H$, and their analytic form is discussed in Ref.~\onlinecite{Sarma1:2012}. Both of these features can be observed in Figure~\ref{Figure_MF_wavefunc_1} where we have plotted the Majorana mode wavefunction across the entire length of the chain, obtained by direct numerical diagonalization of Eq.~\ref{Eq_H_1}. Figure~\ref{Figure_MF_wavefunc_1} shows the spatial extent of the Majorana wavefunctions for two different parameter sets, contrasting their wavefunction overlap.

\begin{figure}[ht]
\centering
\includegraphics[scale=0.22]{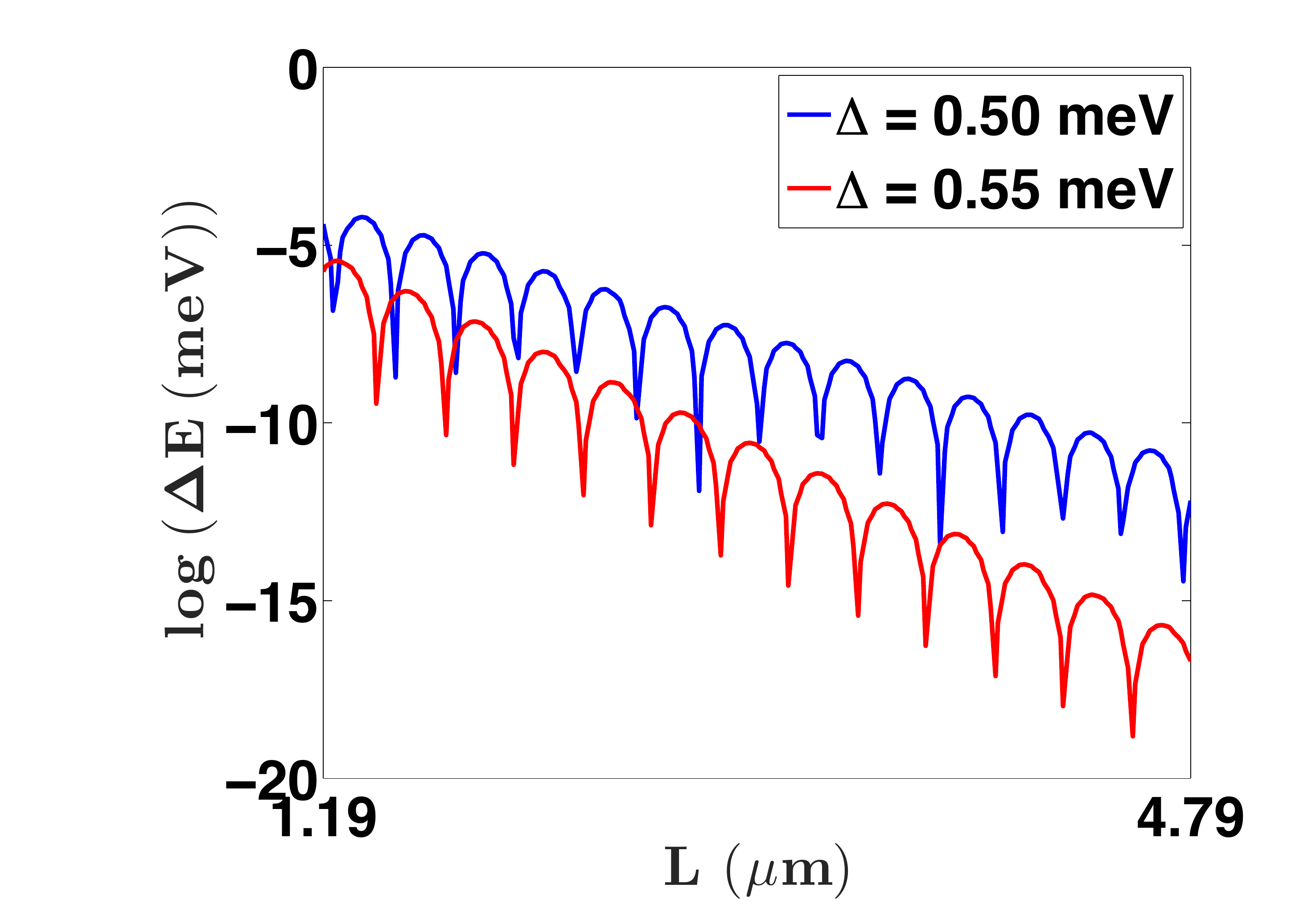}
\includegraphics[scale=0.21]{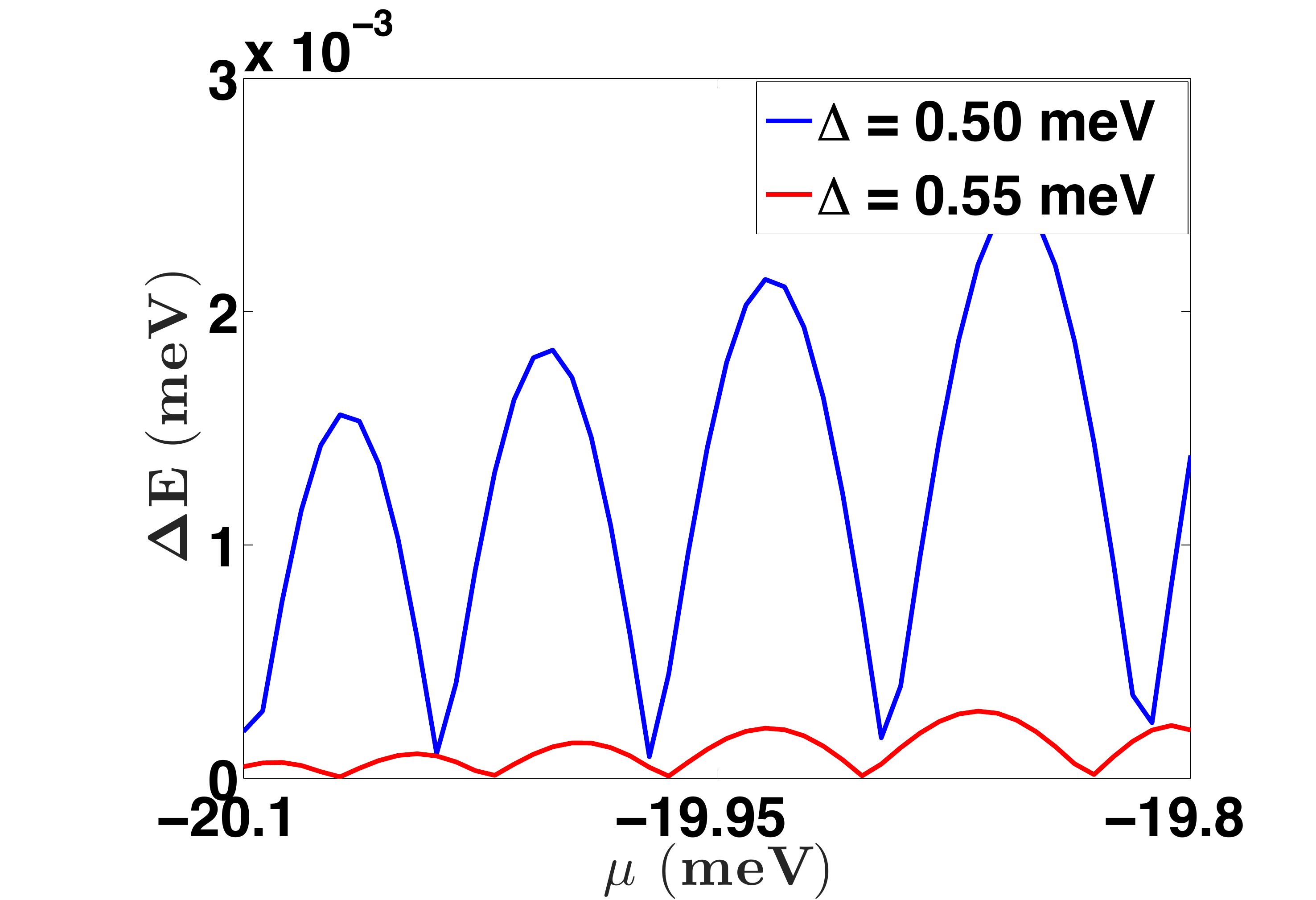}
\caption{\textit{Top panel:} Logarithm of Majorana modes energy splitting $\Delta E$ as a function of the chain length ($L$) for two different values of $\Delta$. As the wire length increases $\Delta E$ decreases exponentially. The value of magnetic field chosen for this plot was $h_x=0.6 meV$. \textit{Bottom panel:} Energy splitting $\Delta E$ (in $meV$) as a function of the chemical potential $\mu$ for two different values of $\Delta$. In both plots the oscillatory behavior of $\Delta E$ is also seen as predicted by Eq.~\ref{Eq_Delta_E_1}. The length of the wire chosen for this plot was $L=1.2 \mu m$.}
\label{Figure_splitting_1}
\end{figure}

\begin{figure}[ht]
\includegraphics[scale=0.14]{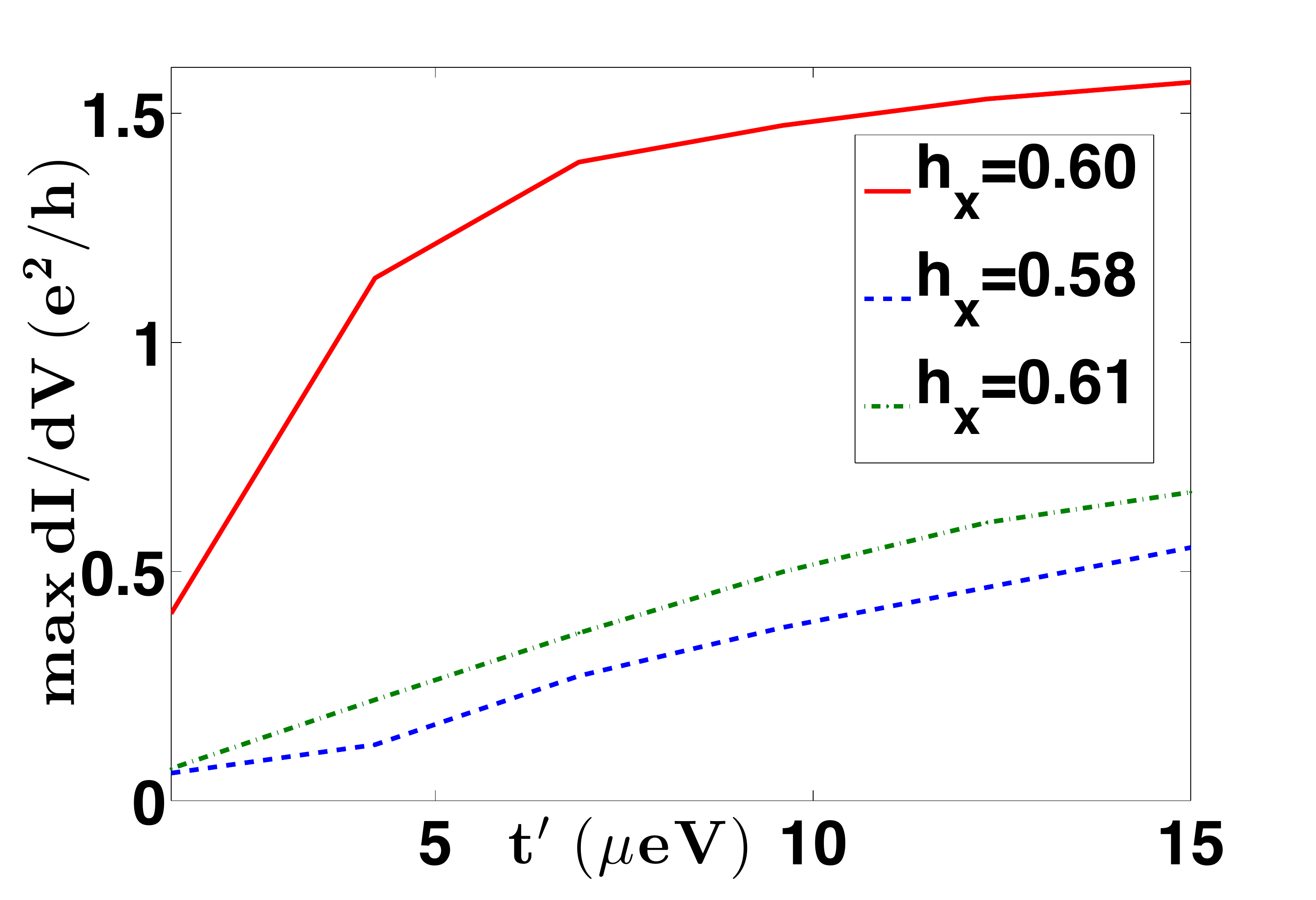}
\includegraphics[scale=0.14]{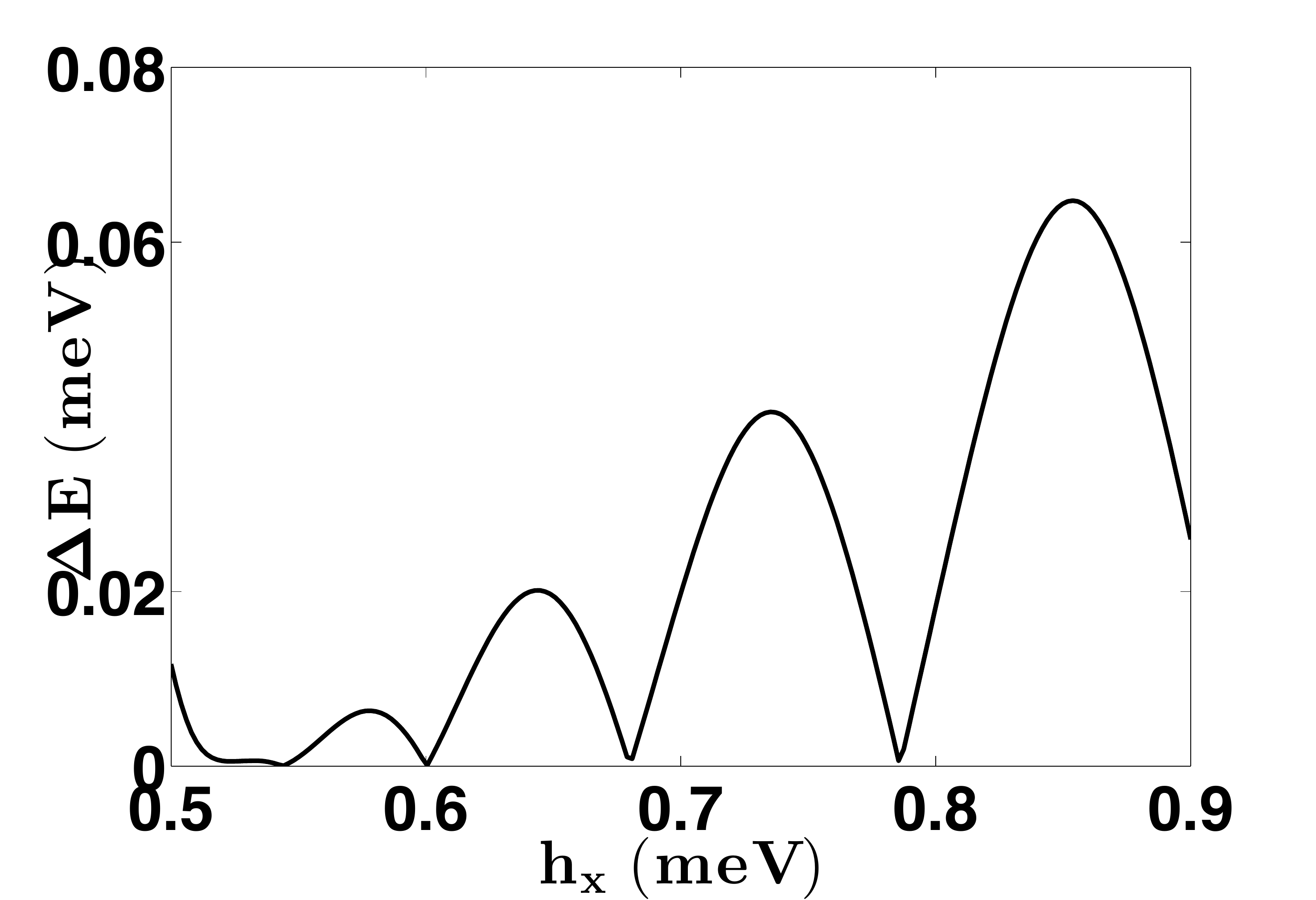}
\caption{(color online) \textit{Left panel:} Maximum peak height of differential conductance $(\text{max }dI/dV)$ as a function of tunneling strength $t'$ in the weak tunneling range $t'\in(1\mu eV,15 \mu eV)$, for three different values of $h_x$. \textit{Right panel:} Majorana mode energy splitting as a function of $h_x$ showing a local minima at $h_x=0.60 meV$. Exactly at $h_x=0.60 meV$, where $\Delta E$ is minimum, the quantized peak height is closer to $G_M$ when compared to values away from the local mimima. We used $\Delta=0.50 meV$ for these plots and $T=0K$. }
\label{dIdV_tunneling_1}
\end{figure}

As a result of Majorana modes hybridization, the zero energy eigenvalues of the Majorana modes are shifted to finite non-zero energies\cite{Cheng:2009, Cheng:2010}. In the limit when $L\gg\zeta$, the splitting energy can be approximately written as~\cite{Sarma1:2012}
\begin{align}
\Delta E \approx \hbar^2 k_{F\text{eff}} \frac{e^{-2L/\zeta}}{m\zeta} \cos(k_{F\text{eff}}L),
\label{Eq_Delta_E_1}
\end{align}
where $m$ is the effective electron mass and $L$ is the length of the nanowire. Eq.~\ref{Eq_Delta_E_1} suggests that $\Delta E$ oscillates as a function of $L$ and $k_{F\text{eff}}$ because of the cosine term. However the amplitude of these oscillations is exponentially suppressed with the wire length $L$, due to the overriding factor $e^{-2L/\zeta}$. Now $k_{F\text{eff}} = k_{F\text{eff}}(h_x,\Delta,\mu,\alpha)$, therefore $\Delta E$ as a function of $\mu$ or $h_x$, should, in principle, show this oscillatory behavior. These features have been highlighted in Figure~\ref{Figure_splitting_1}, which shows $\log(\Delta E)$ as a function of $L$, and $\Delta E$ as a function of the chemical potential $\mu$. Clearly, for higher values of chain length $L$, the energy splitting $\Delta E$ falls exponentially, but one also notes that $\log(\Delta E)$ is not a monotonic function of $\mu$ or $L$, but rather shows an oscillatory behavior as suggested by Eq.~\ref{Eq_Delta_E_1} . Figure~\ref{dIdV_tunneling_1} shows $\Delta E$ as a function of $h_x$ showing similar oscillations. Even though the amplitude of the oscillations in $\Delta E$ decrease exponentially with the length of the chain, for shorter length $L$ (i.e. $L\sim\zeta$) the amplitude of these oscillations can vary over 2-3 orders of magnitude as suggested by the plots in the figures. Therefore in order to minimize $\Delta E$ for smaller values of $L$, a fine-tuning of microscopic parameters like $\Delta$, $\mu$, $h_x$ and $\mu$ is required. 


To show the effect of splitting oscillations due to finite length of the wire, we focus on the weak-tunneling regime (defined in Sec. II) by first choosing $t'\sim 10 \mu eV$. In Figure~\ref{dIdV_tunneling_1} (left panel) we have plotted the maximum peak height of $dI/dV$ as a function of tunneling strength $t'$, for three different values of magnetic field $h_x$. According to Eq.~\ref{Eq_dIdV}, the maximum peak height ($G_M$) is attained in the regime of weak tunneling (see Sec. II) and in an infinite wire. However, for a finite wire length and with weak tunneling, the height of the peak at $\pm \Delta_{\text{lead}}$ is reduced from $G_M$ due to overlap of the Majorana wave-functions. Moreover, in the presence of wave function overlap the peak height is further reduced with reduction of $t'$.

Figure ~\ref{dIdV_tunneling_1} (right panel) shows the energy splitting $\Delta E$ between the two Majorana modes as a function of applied $h_x$, showing an oscillatory dependence of $\Delta E$ on $h_x$. When $\Delta E$ is at a local minima, and thus very close to zero, (for example $h_x$ is fine-tuned at $h_x=0.60 meV$ in Figure~\ref{dIdV_tunneling_1}, corresponding to $\Delta E\sim 10^{-4}$), even a very weak tunneling $t'$ can give rise to a peak height comparable to $G_M$, which otherwise is suppressed by almost an order of magnitude for the same value of $t'$ (see Figure~\ref{dIdV_tunneling_1} left panel). For a concrete comparison we note from Figure~\ref{dIdV_tunneling_1} , that a variation of $h_x$ from $0.58 meV$ to $0.60 meV$, or from $0.61 meV$ to $0.60 meV$ enhances the quantized peak height ($\max dI/dV$) by almost one order of magnitude. Our results suggest that the reduction in the peak height from $G_M$ is a direct consequence of a finite non-zero energy splitting $\Delta E$ between the two Majorana modes. Thus for a finite length of the nanowire, the maximum quantized peak height value can be attained only accidentally even with a SC lead. 

\begin{figure}[ht]
\centering
\includegraphics[scale=0.14]{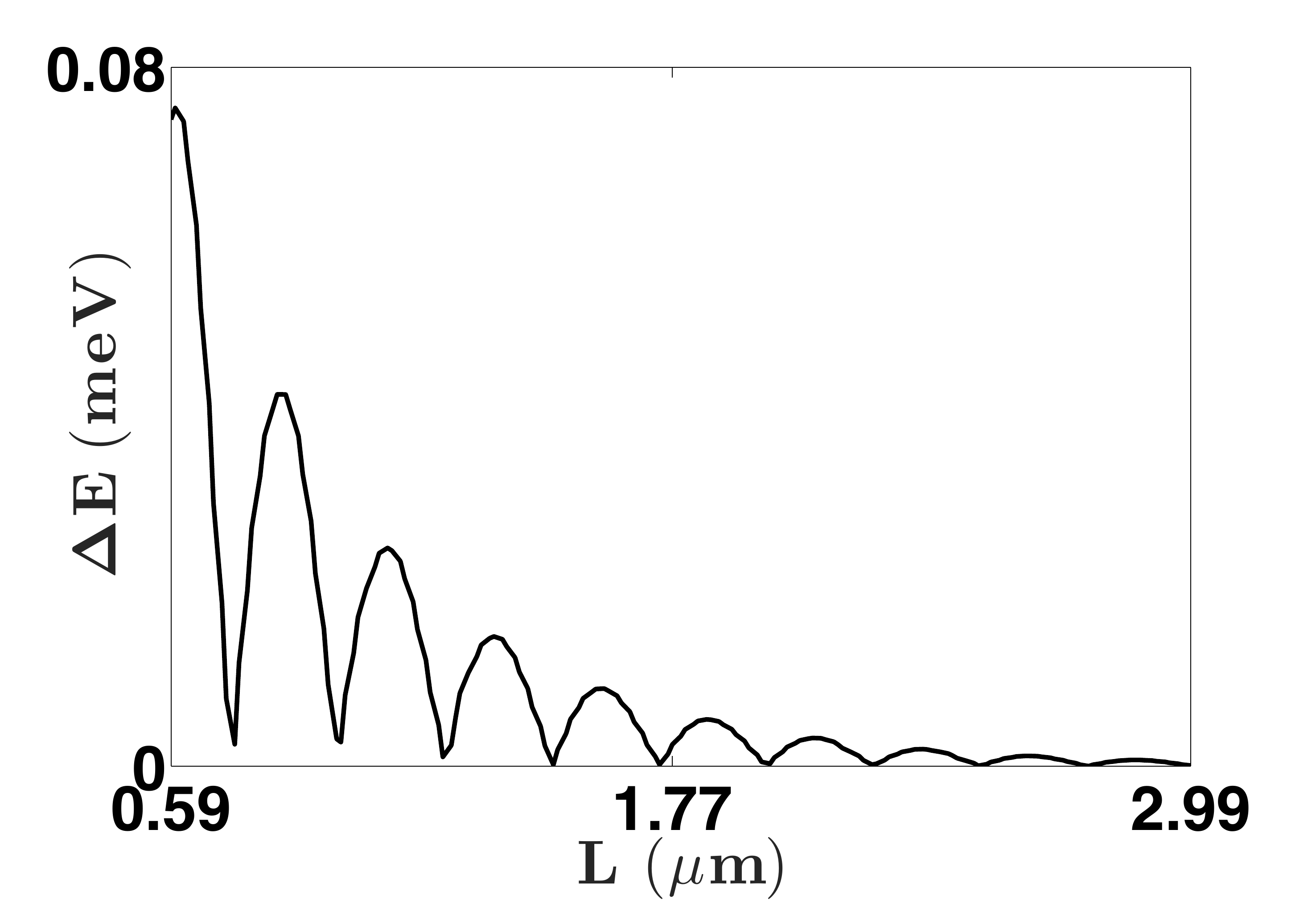}
\includegraphics[scale=0.14]{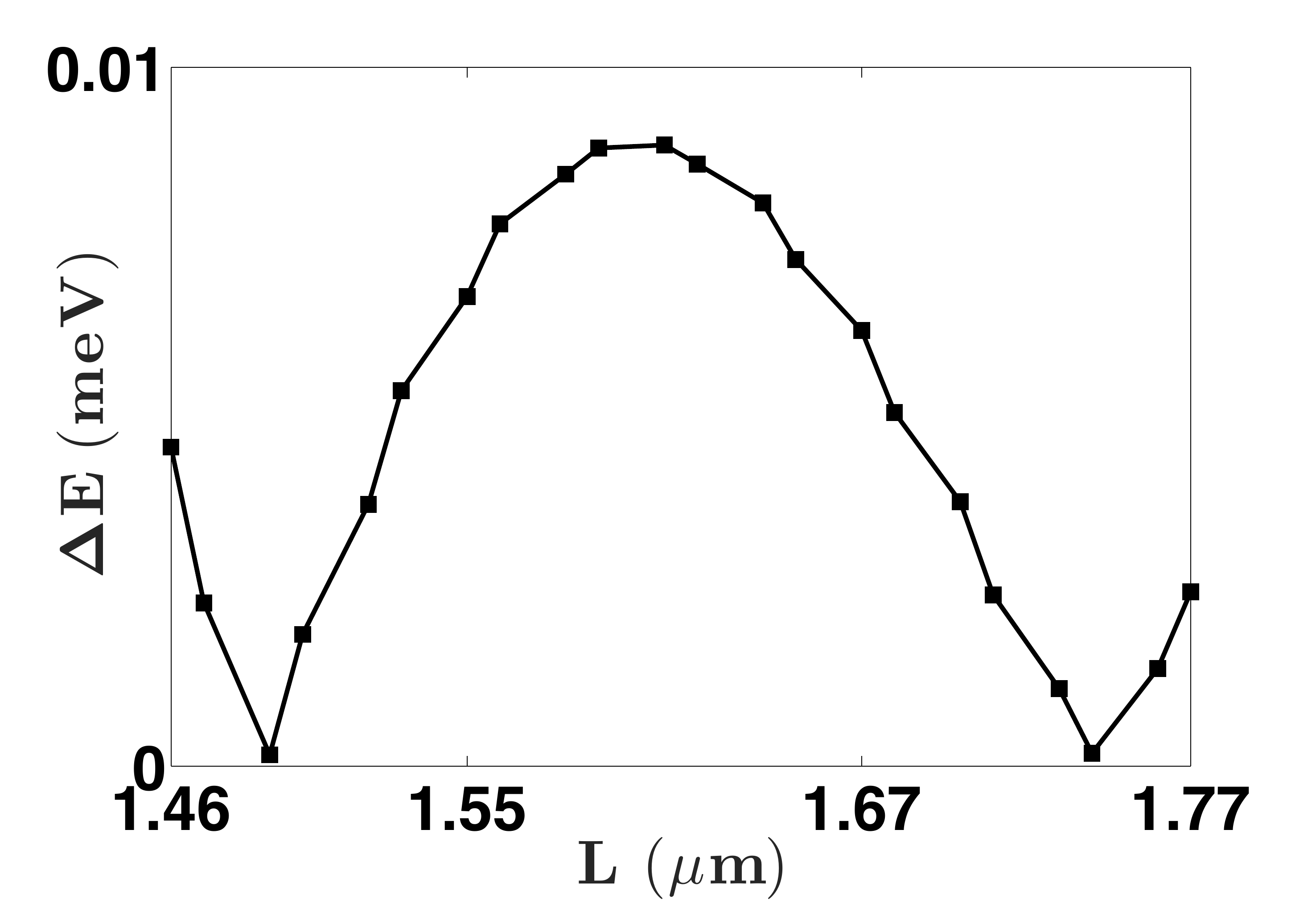}\\
\includegraphics[scale=0.14]{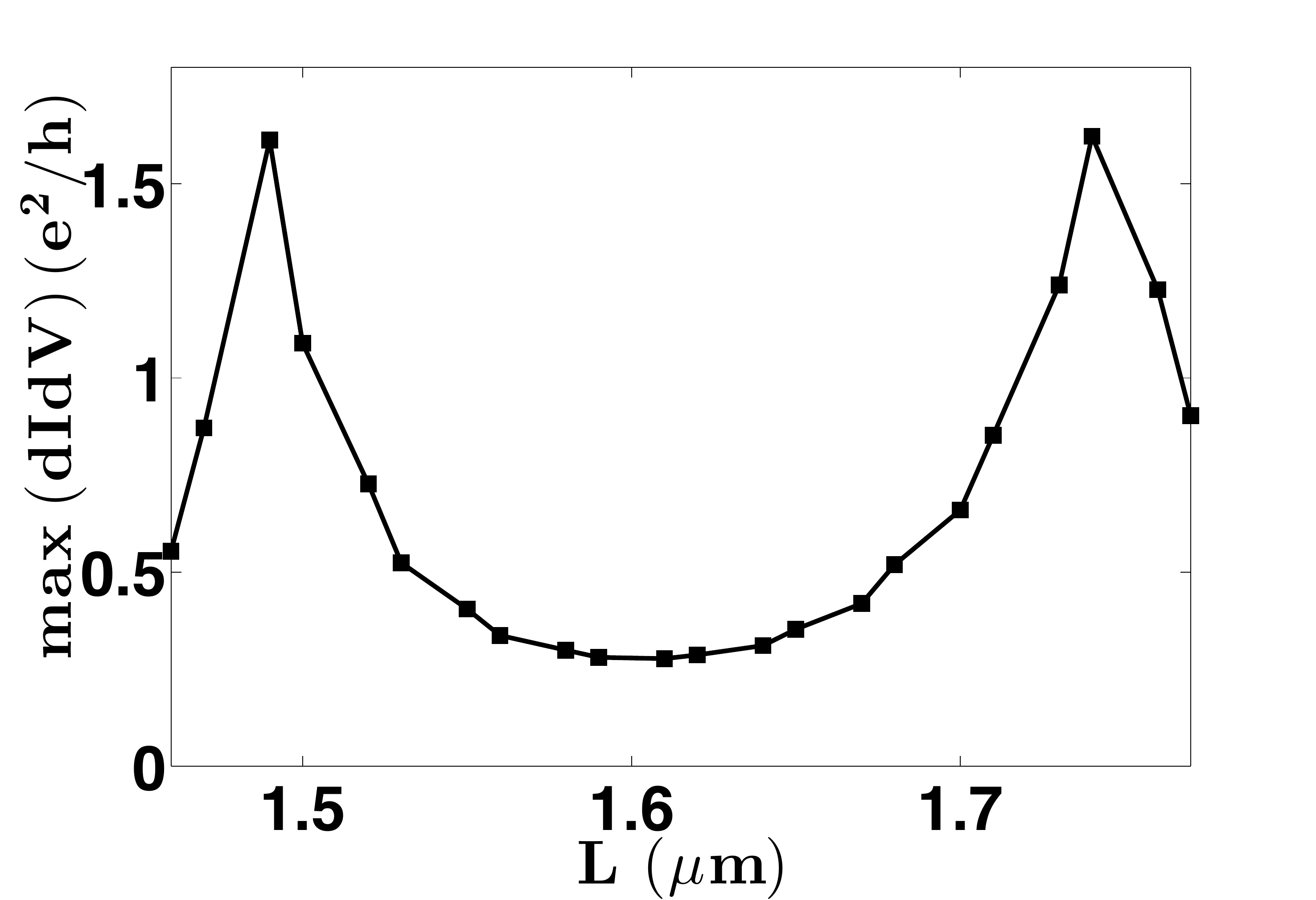}
\includegraphics[scale=0.14]{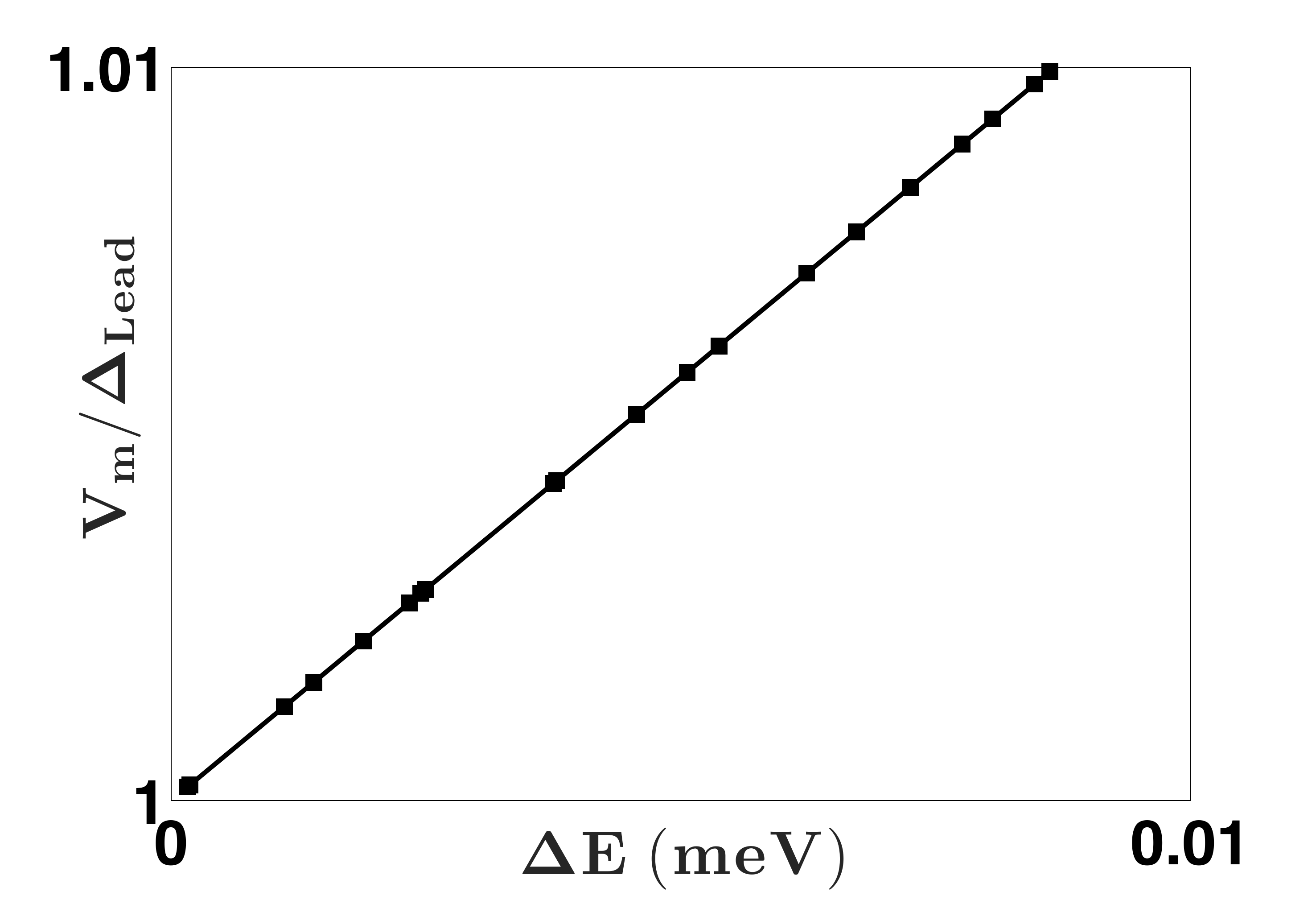}
\caption{\textit{Top panel (left):} Energy splitting $\Delta E$ as a function of chain length $L$. The energy splitting is an oscillatory function of the chain length and is modulated by an overall exponentially decaying function. \textit{Top panel (right):} Energy splitting $\Delta E$ zoomed between chain lengths $L=1.46 \mu m$ and $L=1.77 \mu m$. \textit{Bottom panel (left):} The maximum quantized peak height in the units of $e^2/\hbar$ zoomed in between chain lengths $L=1.46 \mu m$ and $L=1.77 \mu m$. The plots suggest that there is a correspondence between the energy splitting and peak height explicitly showing that the reduction in the peak height is a consequence of the energy splitting between the Majorana modes. The local minima in $\Delta E$ corresponds to a local maxima in the peak height. \textit{Bottom panel (right):} Voltage corresponding to the maxima of the conductance peak $V_{m}/\Delta_{\text{lead}}$ as a function of $\Delta E$. The parameters used for these plot were:  $h_x=0.6 meV$, $\Delta=0.5 meV$, $t'=30 \mu eV$, and $T=0K$.}
\label{Figure_Delta_E_vs_chain_2}
\end{figure}

A similar dependence of the peak height on the chain length $L$ (through the energy splitting $\Delta E$) is presented in Figure~\ref{Figure_Delta_E_vs_chain_2}, where we have plotted, the energy splitting $\Delta E$, and the $dI/dV$ peak height, as a function of chain length $L$ (for a higher value of $t'=50\mu eV$ in this case). The energy splitting is an oscillatory function of the chain length, and is enveloped by an overall exponentially decaying function $e^{-x/\zeta}$. As $L$ is varied from $L=1.60 \mu m$ to $L=1.50 \mu m$, $\Delta E$ changes from $\Delta E\sim 10^{-4}$ to $\Delta E\sim 10^{-2}$, and the peak height reduces from $\max (dI/dV)\sim 1.5e^2/h$ to $\max (dI/dV)\sim 0.5e^2/h$. 

An important feature to be noted from Figure~\ref{Figure_Delta_E_vs_chain_2} and Figure~\ref{dIdV_tunneling_1} is the sensitivity of $\Delta E$ and the $dI/dV$ peak height on experimental parameters, when $L$ is small. This automatically implies a need for fine-tuning the microscopic parameters such as $\mu$, $\Delta$ and so on for smaller values of the chain length $L$, in order to observe a quantized $dI/dV$ peak of the order of magnitude of $G_M$. However such a fine-tuning of various parameters is not possible generically. Thus to observe a quantized peak height $G_M$ for $V=\pm \Delta_{\text{lead}}$ with a SC lead, one requires a long enough wire length $L$, such that the amplitude of splitting oscillations is exponentially suppressed. In shorter wires, the quantization of the peak height is possible, but only when $\Delta E$ is very small which requires a fine-tuning of parameters. 

\begin{figure}[ht]
\centering
\includegraphics[scale=0.14]{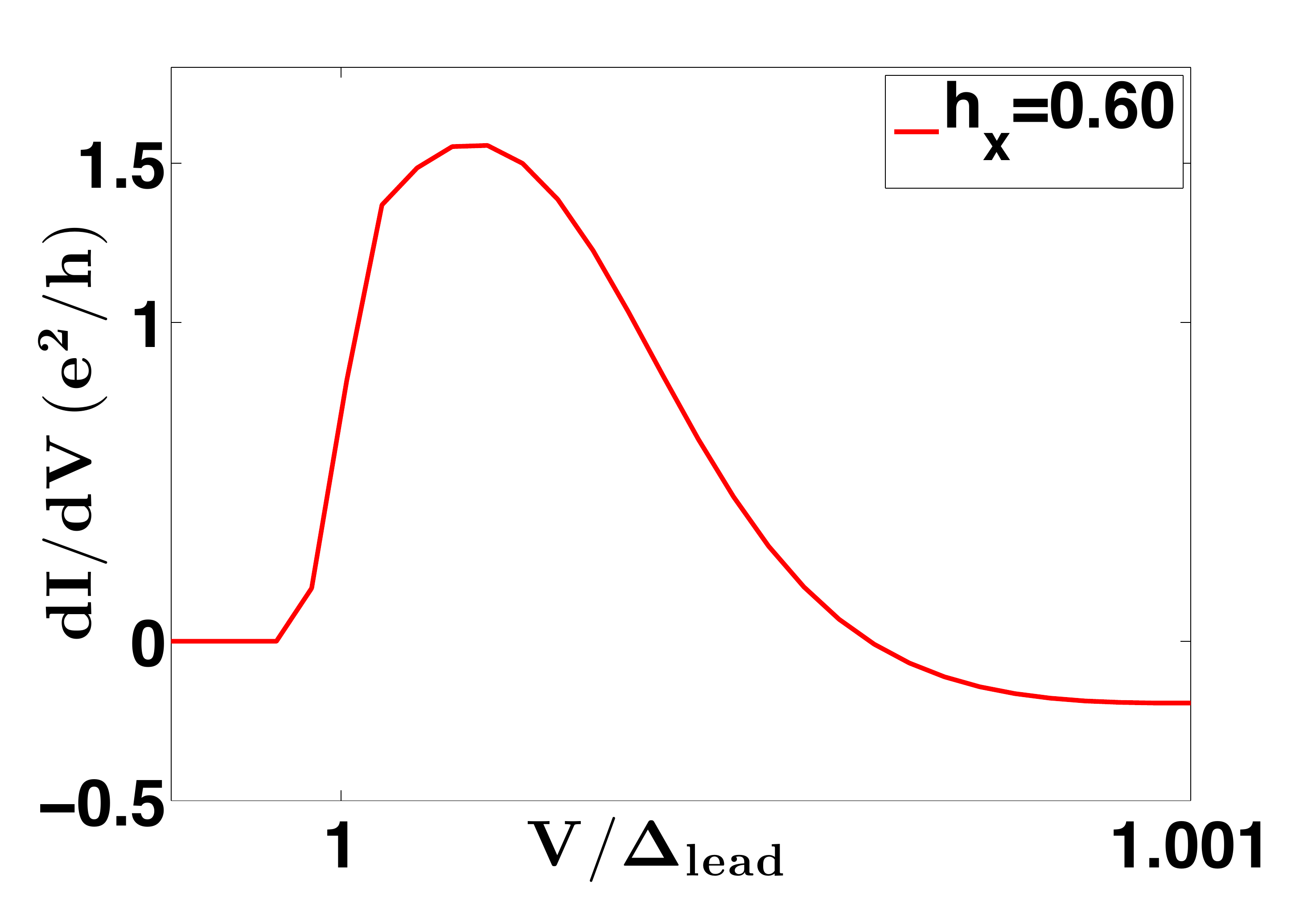}
\includegraphics[scale=0.14]{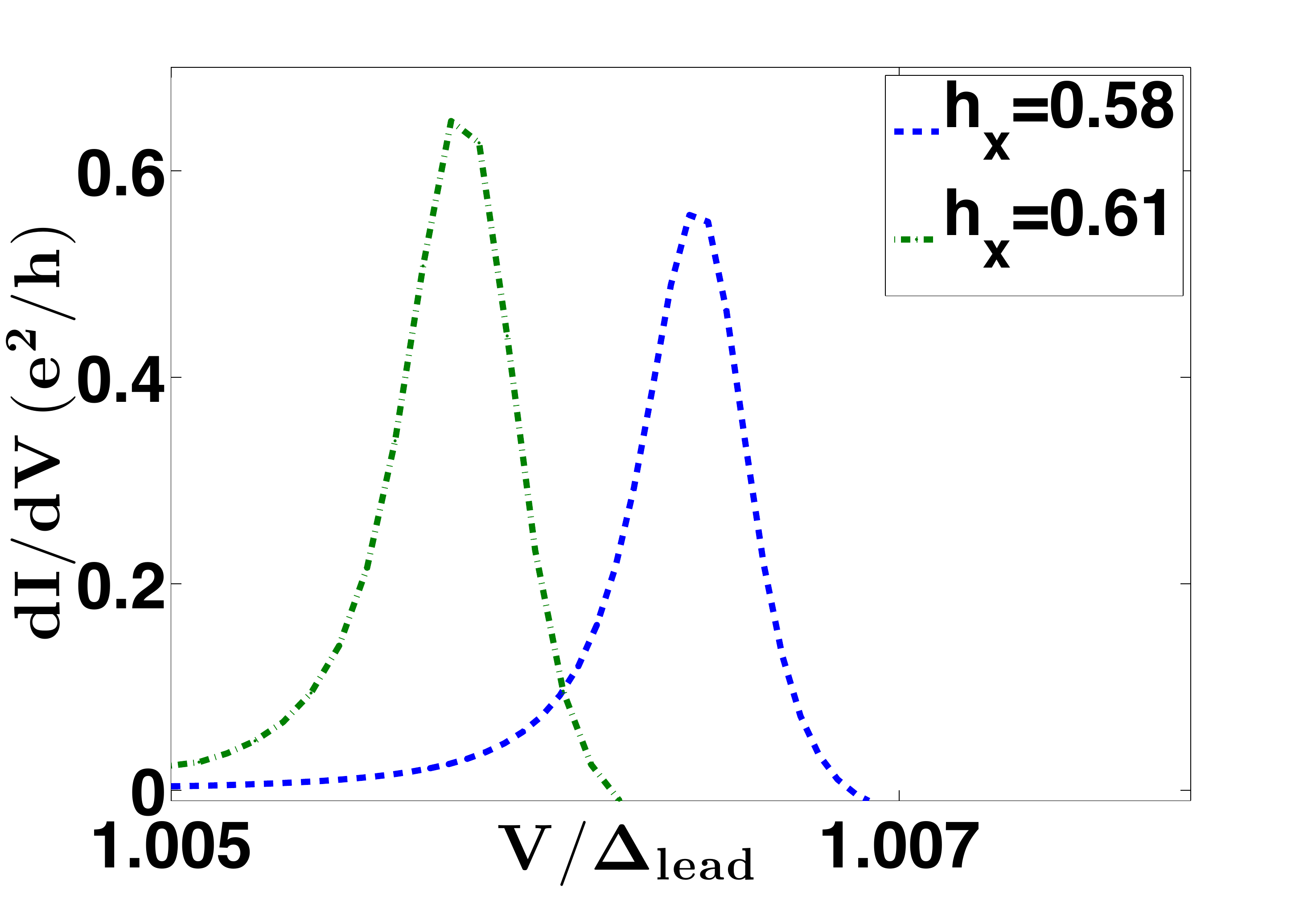}
\includegraphics[scale=0.14]{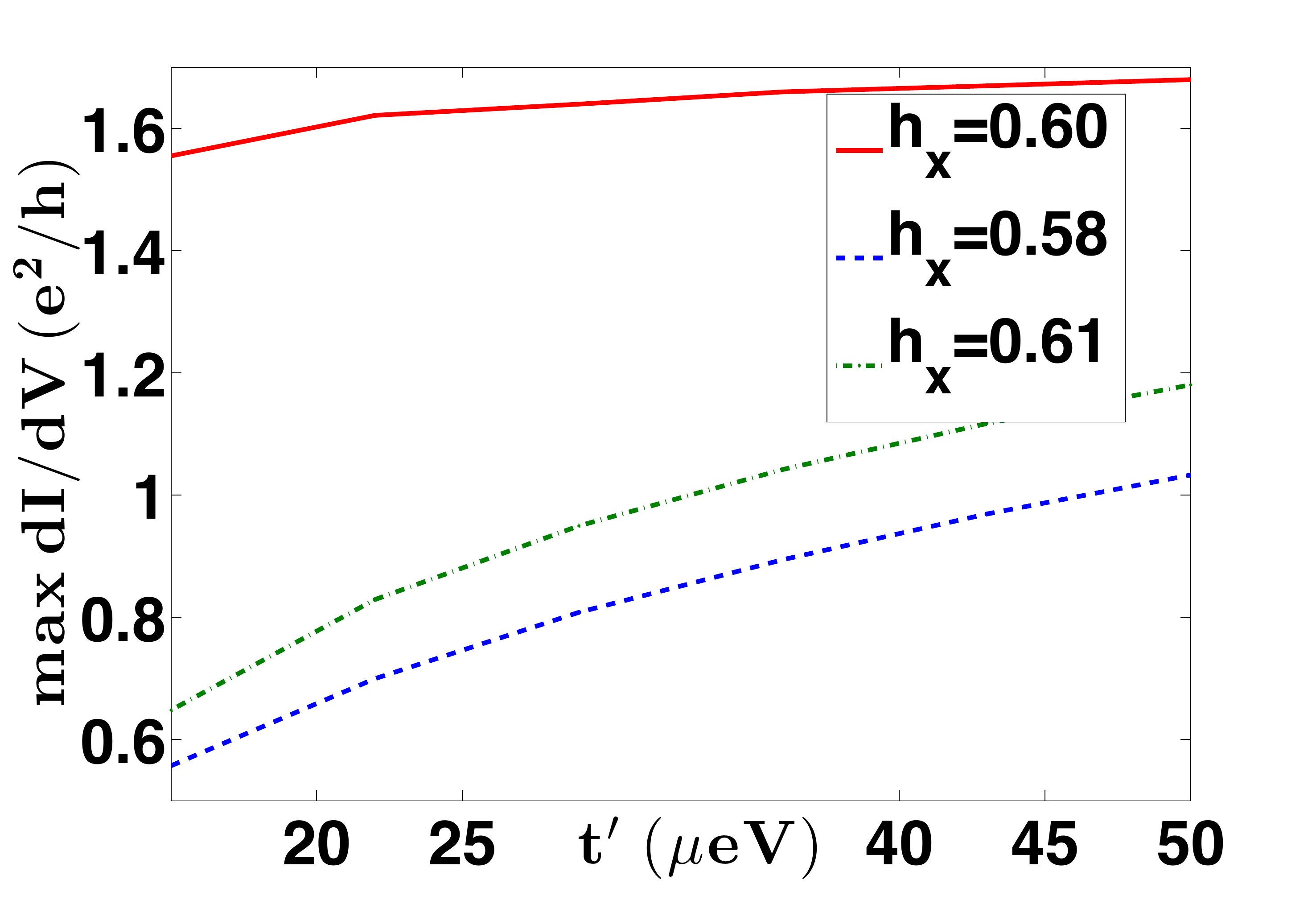}
\includegraphics[scale=0.14]{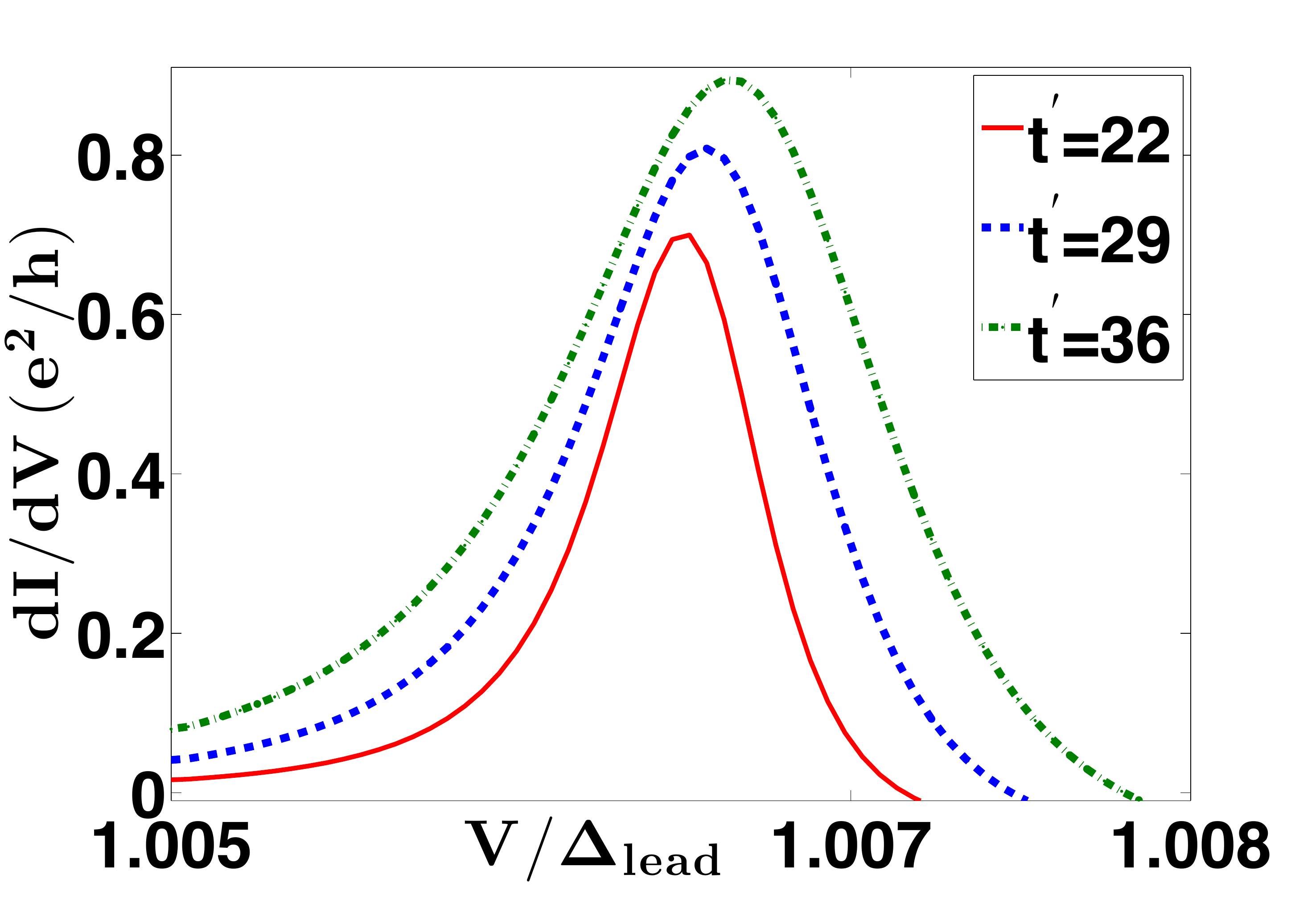}
\caption{(color online) \textit{Top panel:} $dI/dV$ lineshape for three different values of $h_x$ at $t'=15 \mu eV$. \textit{Bottom panel (left):}  Maximum peak height of differential conductance $(\text{max }dI/dV)$ as a function of tunneling strength $t'$ in the range $t'\in(15\mu eV,50 \mu eV)$, for three different values of $h_x$.  \textit{Bottom panel (right):} $dI/dV$ profile for $h_x=0.58 meV$ for three different values of $t'$ (in the units of $\mu eV$). The value of the SC gap chosen in all these plots was $\Delta=0.5 meV$ (the same as used in Figure~\ref{dIdV_tunneling_1} and Table I). The plots suggest that even when $t'$ is not very weak (compared to $t'$ Figure~\ref{dIdV_tunneling_1}), there is a broadening of the lineshape accompanied by an offset of threshold voltage from $V=\Delta_{\text{lead}}$, when $\Delta E$ is significant. Further $T=0K$.}
\label{Figure_dIdV_vs_chain_3}
\end{figure}

In the limit of weak tunneling $t'$, the zero-bias conductance peak of the Majorana zero-energy mode in normal metal tunneling junction is split into two symmetric peaks around $V=0$ as a consequence of hybridization of the Majorana modes due to finite wire lengths. The splitting gap is typically of the order of $\Delta E$. Therefore, strictly speaking, at $T=0K$, a zero bias peak at exactly $V=0$ should not be ideally observed, and rather two symmetric peaks around $V=0$ should appear for a short wire. However a small finite temperature broadens the lineshape, and the two split peaks then appear as a single symmetric peak centered at $V=0$. These features have been numerically studied for a metal-SC tunneling contact~\cite{Lin:2012,Dumitrescu1:2015}. In the present case, where we consider a SC lead, the conductance peak at $V=\pm\Delta_{lead}$ does not split as a result of non-zero $\Delta E$, unlike in the case of metallic lead. Instead the threshold voltage $V_{th}$, where the $dI/dV$ conductance peak exhibits a sharp rise from zero to $G_M$, shifts (by approximately$\sim\Delta E$) from $V=\pm\Delta_{\text{lead}}$. We present this feature in Figure~\ref{Figure_Delta_E_vs_chain_2} where we have plotted $V_{m}/\Delta_{\text{lead}}$ increasing monotonically as a function of energy splitting $\Delta E$ between the two Majorana modes, where $V_m$ is the voltage at which $dI/dV$ is maximum. Note that due to lineshape broadening, a sharply rising $dI/dV$ peak will not be observed (see Figure~\ref{Figure_dIdV_vs_chain_3} and the discussion below), and therefore we have plotted $V_{m}/\Delta_{\text{lead}}$ as the threshold voltage $V_{th}$ is not sharply defined. Also, since $I(-V)=-I(V)$, the $dI/dV$ response for a negative bias voltage is symmetric. 

\begin{figure}[ht]
\centering
\includegraphics[scale=0.14]{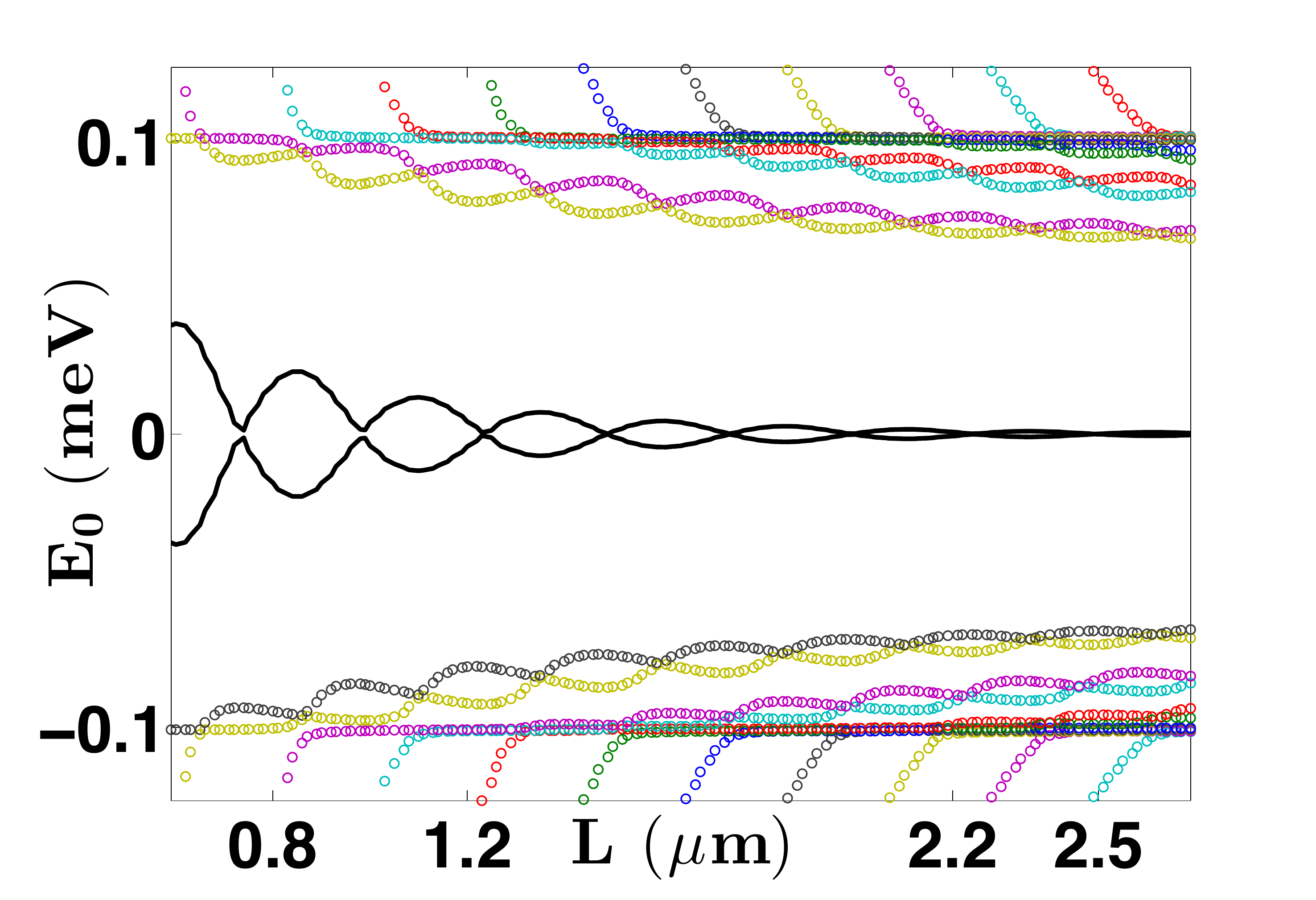}
\includegraphics[scale=0.14]{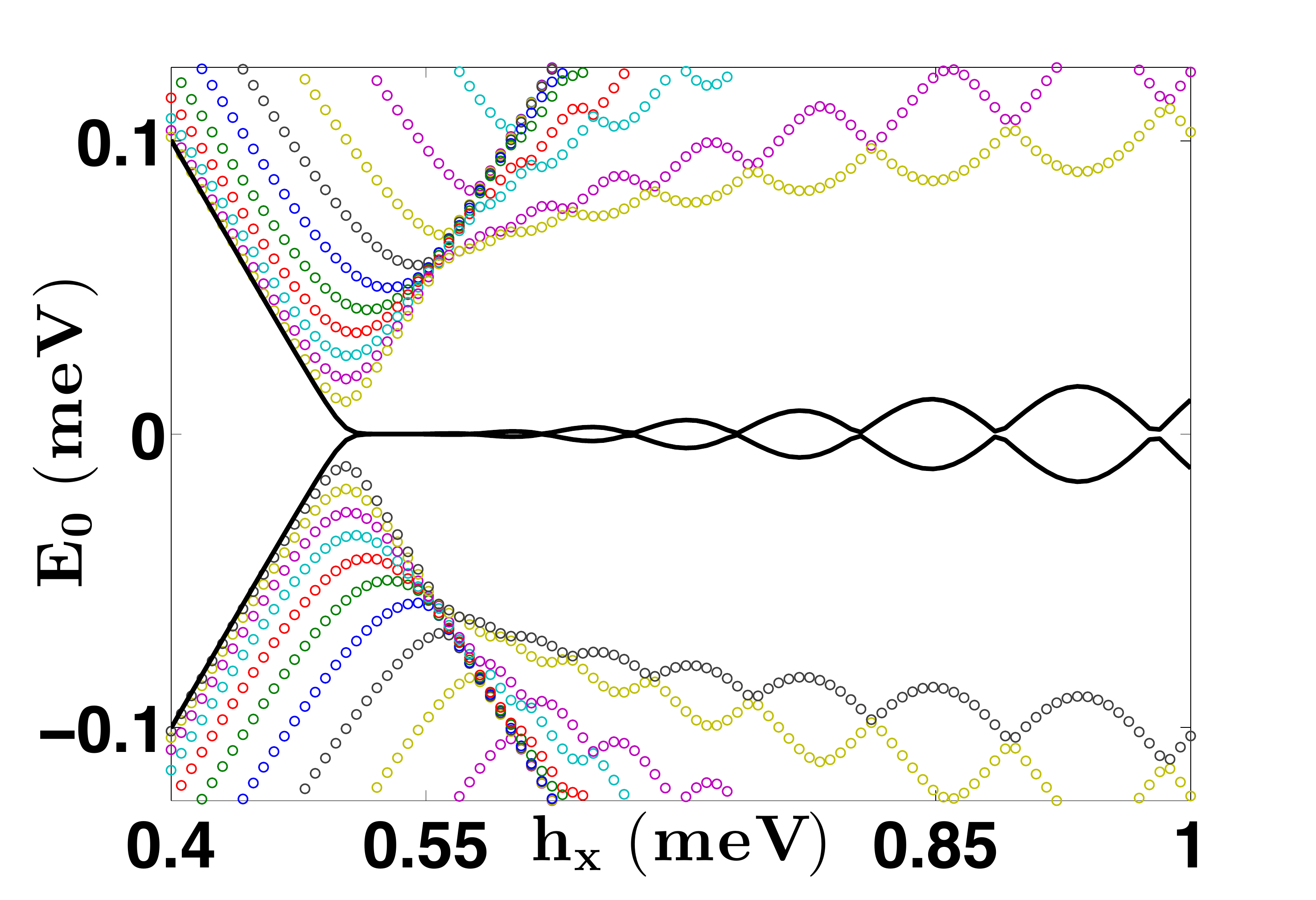}
\caption{(color online) \textit{Left panel:} Quasiparticle spectrum of the TS phase of the 1D nanowire (in color), along with the zero energy Majorana bound state $(E_0)$ (in black) as a function of wire length $L$, for $h_x=0.6 meV$. The Majorana mode which occurs at exactly zero energy when $L\rightarrow\infty$ is adiabatically connected to the quasi-Majorana mode in short wires. \textit{Right panel:} Quasiparticle spectrum along with the zero energy Majorana bound state $(E_0)$ as a function of $h_x$ for a wire with length $L=1.2 \mu m$, also displaying a topological phase transition at $h_x=0.5 meV$. The value of SC gap chosen for these plots was $\Delta=0.5 meV$.}
\label{Figure_E_0_vs_L}
\end{figure}
\begin{figure}[ht]
\includegraphics[scale=0.27]{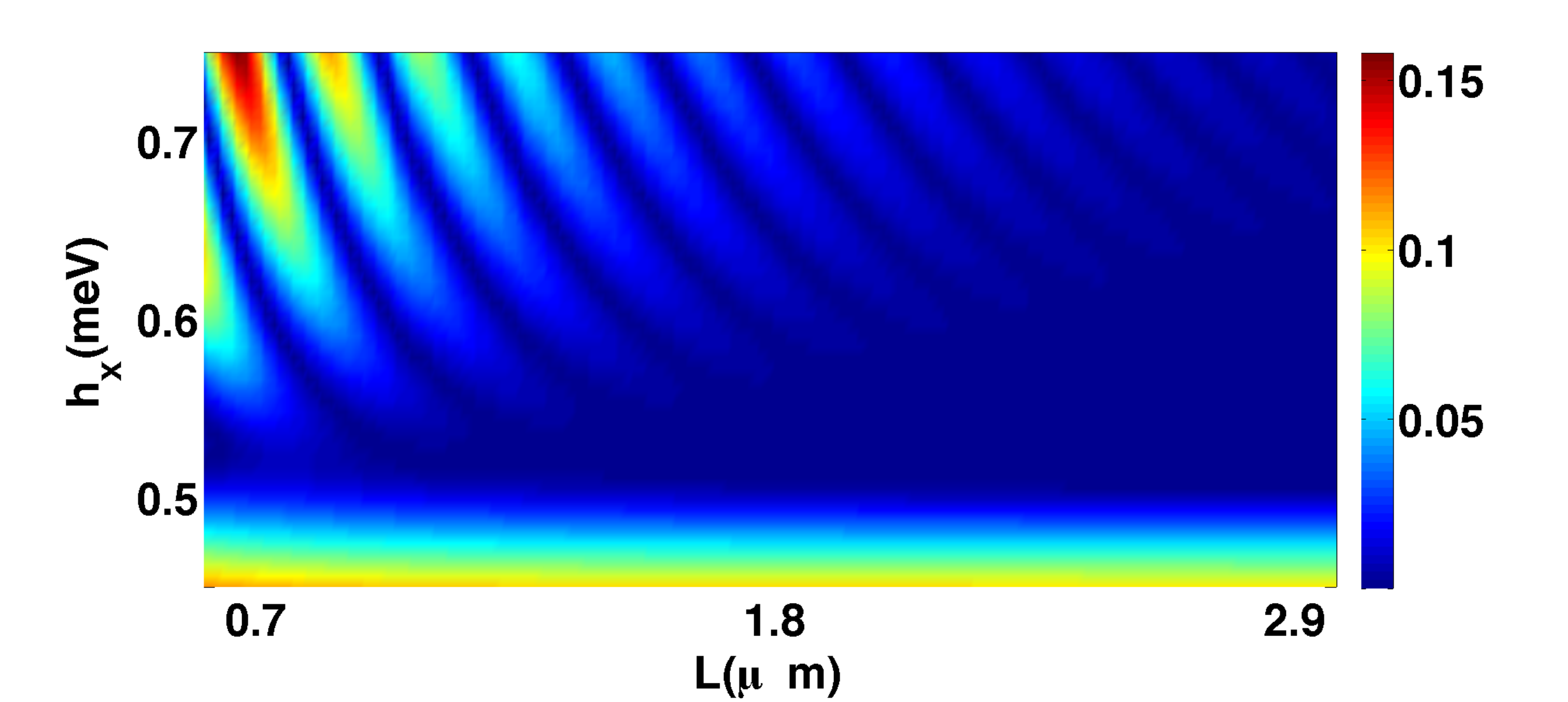}
\caption{(color online) Majorana splitting energy $\Delta E$ as a function of both $L$ and $h_x$, showing oscillations for small $L$ values, for a wide range of $h_x$ above the topological phase transition at $h_x=0.5 meV$. The amplitude of oscillations in $\Delta E$ is subsequently suppressed for wires with longer length. The value of SC gap chosen for these plots was $\Delta=0.5 meV$, and the colorbar on the right is in the units of $meV$.}
\label{Figure_H_x_vs_L}
\end{figure}
\begin{figure}[ht]
\includegraphics[scale=0.45]{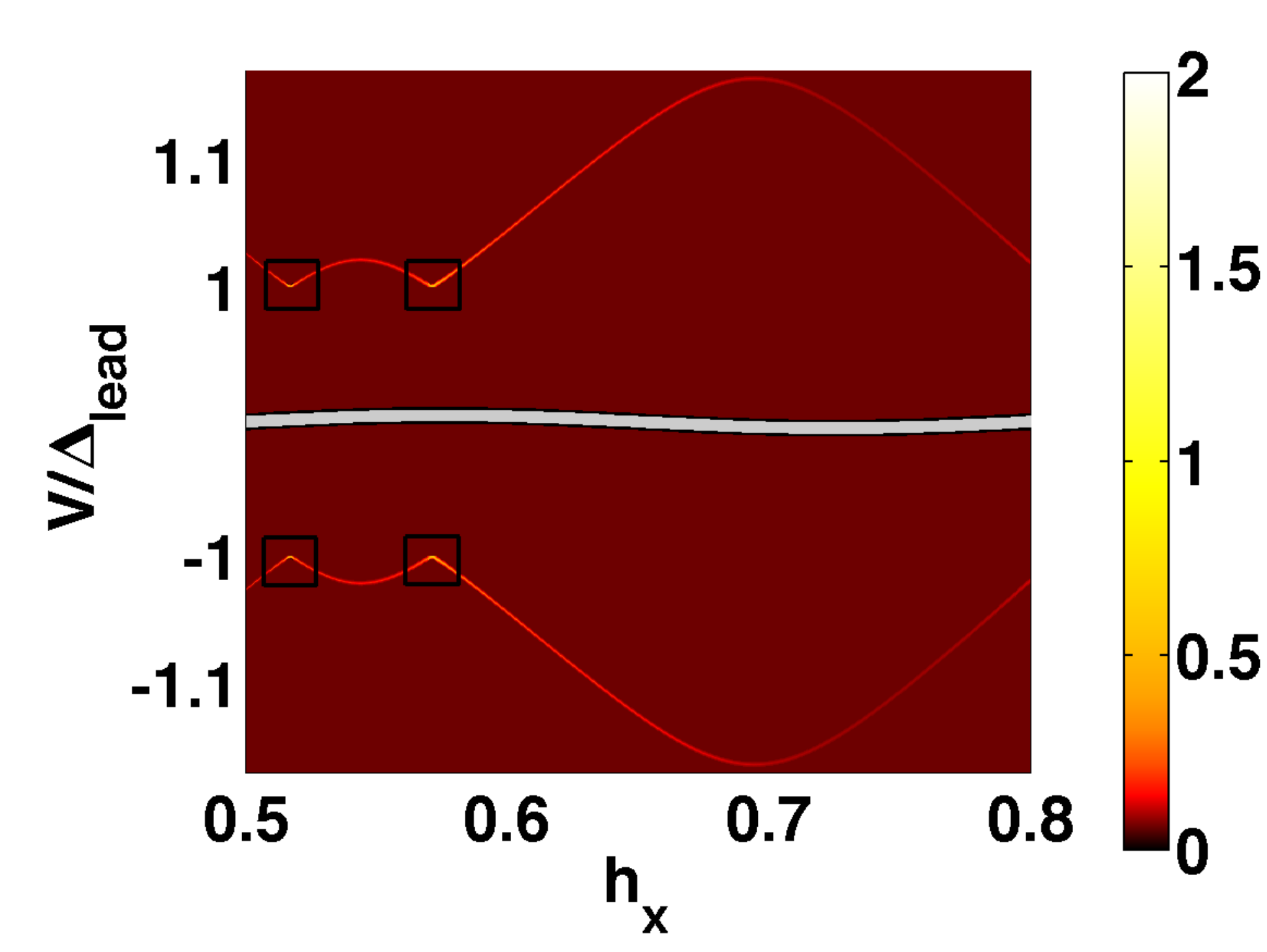}
\includegraphics[scale=0.25]{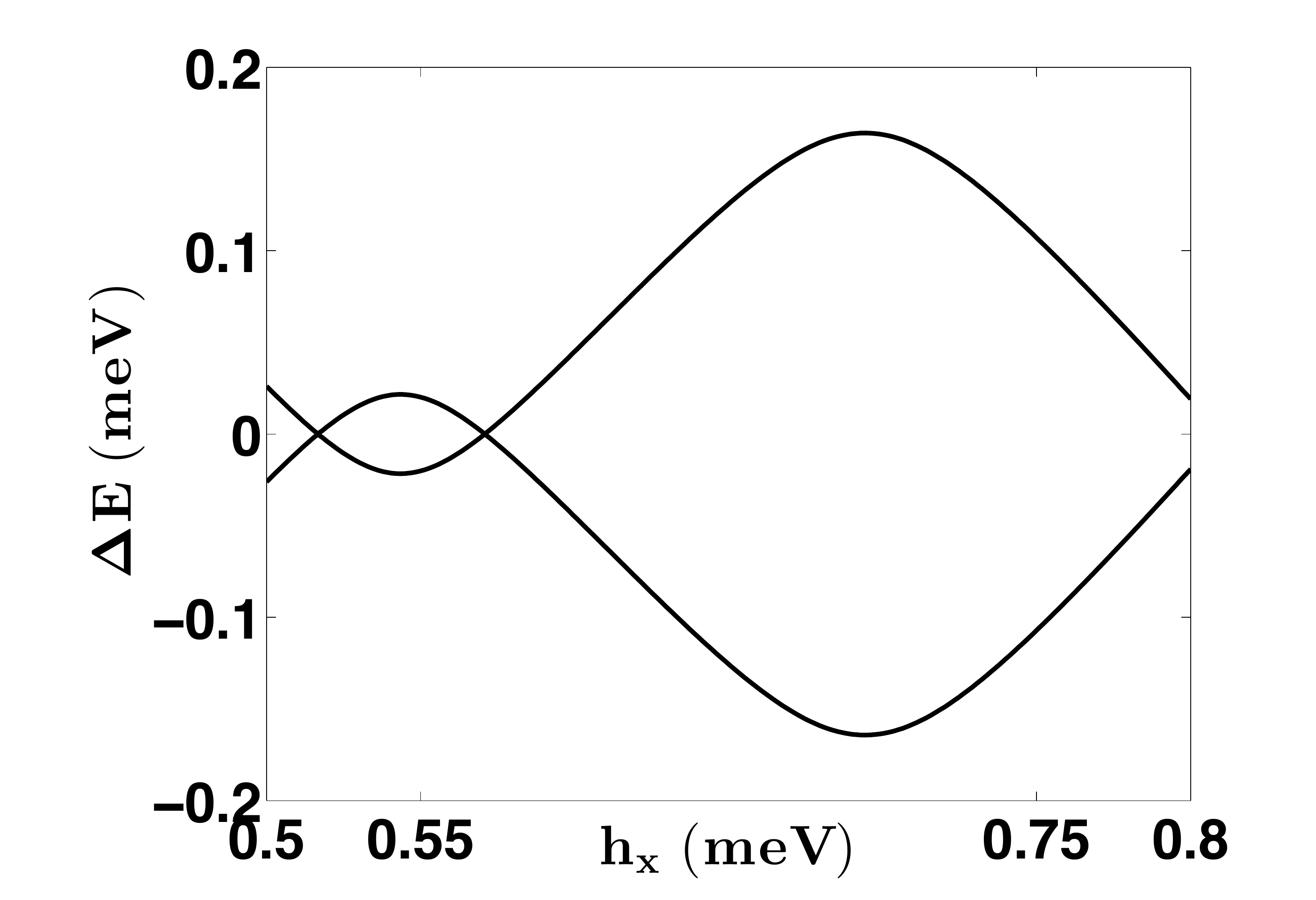}
\caption{(color online) \textit{Top panel:} Majorana $dI/dV$ profile for a short wire of length $L=0.54\mu m$, showing oscillations about $V=\pm\Delta_{\text{lead}}$ as a function of magnetic field $h_x$ (in $meV$). The colorbar on the right is in the unit of $e^2/h$, the value of SC gap chosen was $\Delta=0.5 meV$, and $t'=50 \mu eV$. The four square boxes highlight the regions where the quantized peak height of magnitude $G_M=(4-\pi)e^2/h$ should be observed. \textit{Bottom panel:} The energy splitting $\Delta E$ (in $meV$) as a function of $h_x$ for the same system, showing oscillations about zero energy. In an experiment done with a normal metallic lead, the oscillations will be about zero bias voltage in a similar fashion as displayed in this plot.}
\label{Figure_colorplot}
\end{figure}

Figure~\ref{Figure_dIdV_vs_chain_3} shows results for the peak height and the lineshape for a larger value of $t'$, but for the same three different $h_x$ values, and for the same set of parameter values as used in Figure~\ref{dIdV_tunneling_1} and Figure~\ref{Figure_Delta_E_vs_chain_2}. Our conclusions on the dependence of the lineshape and the peak height on $\Delta E$ do not change qualitatively. However making the barrier more transparent, i.e. increasing $t'$, also results in a corresponding increase in the overall peak height. Further, we also illustrate in Figure~\ref{Figure_dIdV_vs_chain_3} the broadening of the lineshape as a consequence of non-zero energy splitting $\Delta E$ in a finite wire. In this broadened peak, there is no sharply defined threshold voltage $V_{th}$, where the $dI/dV$ conductance quickly rises from zero. Hence, as a result of splitting of the Majorana zero energy modes, it is not just the peak height which is suppressed, but the overall lineshape is also modified, and thus no longer resembles its analytic form as shown in Fig~\ref{dIdV_temperature} (which was valid when $\Delta E\rightarrow 0$). This feature can be contrasted with the effect of finite temperature on the lineshape as shown in Figure~\ref{dIdV_temperature}. In Figure~\ref{dIdV_temperature}, with temperature one observed just an overall suppression of the height of the lineshape, with a sharply rising peak at $V=\Delta_{\text{lead}}$, and the asymmetry of the peak largely intact. 

\subsection{Experimental implications}
Having discussed important aspects of the $dI/dV$ profile, we now discuss the experimental implications of our work. In Figure~\ref{Figure_E_0_vs_L}, we have plotted the quasiparticle spectrum of the TS phase of Hamiltonian $H$ given in Eq.~\ref{Eq_H_1}. Similarly Figure~\ref{Figure_H_x_vs_L} shows Majorana splitting energy $\Delta E$ as a function of both $L$ and $h_x$ together in a color-plot, displaying clear periodic oscillations in $\Delta E$ for small $L$ values. We note from Figure~\ref{Figure_E_0_vs_L} that for short wires, the zero energy Majorana modes bifurcate into finite energies, with periodic zero-energy crossings. We can therefore term them as `quasi-Majorana' modes, which are remnants of the Majorana physics in idealized situations~\cite{Tudor:2013}. PH symmetry always ensures that the overall energy spectrum has a vanishing sum of the energy eigenvalues, resulting in a symmetric spectrum about $E=0$. It is also straightforward to note that these periodic zero-energy crossings of the quasi-Majorana eigenmodes are related to the splitting energy $\Delta E\rightarrow 0$ discussed earlier. Even though a perfect zero energy mode can occur only in the thermodynamic limit $L\rightarrow\infty$, this zero-energy mode is adiabatically connected with the quasi-Majorana mode as shown in Figure~\ref{Figure_E_0_vs_L}. Such an adiabatic connection is not exhibited by any other trivial zero-energy mode. For example, an accidental zero energy Andreev bound state may occur in a short wire, but in long wires these states will be characterized by a finite energy gap, while the energy of the Majorana mode will vanish~\cite{Tudor:2013}. This adiabatic connection can therefore have important experimental implications.
Figure~\ref{Figure_colorplot} shows the $dI/dV$ profile for a Majorana mode in a short wire of length $0.54\mu m$, showing oscillations about $V=+\Delta_{\text{lead}}$ and $V=-\Delta_{\text{lead}}$ as a function of the applied magnetic field $h_x$. Exactly at $V=\pm\Delta_{\text{lead}}$, the quantization of the Majorana peak is attained at isolated values of $h_x$. However, for a broad range of $h_x$ (though within topological regime), a tunneling experiment performed on Majorana nanowires using SC leads should be able to observe similar oscillations about $V=\pm\Delta_{\text{lead}} $, which is directly connected to the adiabaticity of the `quasi' Majorana mode. This is to be contrasted with the case when normal metallic leads are used. The splitting energy then results in oscillations of tunneling conductance about the zero-bias voltage instead of $V=\pm \Delta_{\text{lead}}$. Therefore even with a SC lead, and a low enough temperatures, for the experimentally relevant finite length wires the quantization of Majorana peak height could be hard to obtain. In this case the signature of the MFs would be splitting oscillations of the quasi-Majorana modes as a function of the magnetic field, but around $V=\pm\Delta_{\text{lead}}$, rather than around $V=0$ as in SC-metallic lead tunneling conductance experiments. 

\section{Conclusions}

We discuss the $dI/dV$ spectra using a SC lead of a spin-orbit coupled SM-SC heterostructure nanowire, a system which has been extensively studied both theoretically and experimentally using a normal metallic lead. We consider different set of physical parameters including temperature, tunneling strength at the junction, wire length, magnetic field, and induced SC pairing potential in the nanowire, and find that in a finite wire the Majorana splitting energy $\Delta E$, which shows an oscillatory dependence on these parameters remains responsible for the $dI/dV$ peak broadening, even when the thermal effects are suppressed by SC gap in the lead. Our numerical results explicitly map the oscillations in $\Delta E$, inversely, to oscillations in the peak height. We find that this effect is quantitatively significant in short wires ($L\sim\zeta$), as $\Delta E\sim e^{-\zeta/L}$, and in a less transparent barrier, where a very small variation in $L$ can result in the reduction of the peak height by almost an order of magnitude. In longer wires, since the amplitude of these oscillations falls exponentially, the variation in peak height will be insensitive to  variations of the microscopic parameters, eliminating the need of a fine-tuned system in order to observe the  quantized height of the Majorana peak. 

Secondly, with the use of a SC lead in a short wire, we find that, apart from the broadening of the peak height due to overlapping MF wave functions, the threshold voltages ($V_{th}=\pm\Delta_{\text{lead}}$) where the Majorana $dI/dV$ peaks arise are also shifted by approximately $\Delta E$. This is to be contrasted with the splitting of ZBP around $V=0$ in tunneling conductance experiments using a metallic lead. The splitting of the ZBP in the present case of a SC lead shifts the threshold voltage $V_{th}\rightarrow V_{th}\pm\Delta_{\text{lead}}\pm\Delta_E$. 


Finally, we have also illustrated a distinguishing feature in the conductance lineshape between thermal broadening and energy splitting. When $T\neq 0, \Delta E\rightarrow 0$, thermal effects lead to an overall suppression of the peak height without significantly altering its lineshape. The $dI/dV$ peak in this case sharply rises from zero at $V=V_{th}=\pm\Delta_{\text{lead}}$, and is asymmetric  about $V_{th}$. When $T\rightarrow 0, \Delta E\neq 0$, the lineshape is broadened and appears symmetric about $V_m$ (where $V_m>V_{th}$ is the bias voltage at which $dI/dV$ is maximum). Our main conclusion in this work is that in a finite length SM wire the overlap of the wavefunctions of the MFs for the two ends remains responsible for the broadening of the Majorana peaks, even when the thermal effects are suppressed by a SC lead. In this case the signatures of Majorana fermions with a SC lead are oscillations of quasi-Majorana peaks about bias $V=\pm\Delta_{\text{lead}}$, in contrast to the case of metallic leads where such oscillations are about zero bias. Our results will be useful for
analysis of future experiments on SM-SC heterostructures using SC leads. In our work we have not included effects of interaction between the Majorana modes. Even though Majorana’s do not carry any charge, they can have an effective long interaction through the even-odd electron number dependence of the superconducting ground state~\cite{Heck:2011}. This might further contribute to the splitting energy, which is a topic of future investigation.  

\textit{Acknowledgment}: We thank J. D. Sau for useful discussions. This work is supported by AFOSR (Grant No. FA9550-
13-1-0045).


\end{document}